\documentclass{article}
\usepackage{amsmath,dsfont}
\usepackage{authblk,booktabs}
\usepackage[round]{natbib}
\usepackage{graphicx}
\usepackage[left=1.3in,top=1.25in,right=1.3in,bottom=1.25in]{geometry}
\usepackage{setspace}
\usepackage{amssymb}
\usepackage{paralist}
\usepackage{hyperref}
\usepackage{cleveref}

\DeclareMathOperator{\ddt}{\dfrac{d}{d t}}

\newcommand{\pardif}[2]{\ensuremath{\dfrac{\partial #1}{\partial\,#2}}}
\newcommand{\spardif}[2]{\ensuremath{\frac{\partial #1}{\partial\,#2}}}

\begin{document}

\title{Covariant Lyapunov Vectors of a Quasi-geostrophic Baroclinic Model}
\author[1,2]{Sebastian Schubert}
\author[2,3]{Valerio Lucarini}
\affil[1]{IMPRS - ESM, MPI f. Meteorology, University Of Hamburg, Hamburg, Germany, Email: sebastian.schubert@mpimet.mpg.de}
\affil[2]{Meteorological Institute, CEN, University Of Hamburg, Hamburg, Germany}
\affil[3]{Department of Mathematics and Statistics, University of Reading, Reading, United Kingdom}

\date{First Draft: October 2014\\
Second Draft: March 2015\\
Third Draft: June 2015\\ This Draft: August 2015\\ }
\maketitle
\singlespacing
\begin{abstract}
The classical approach for studying atmospheric variability is based on defining a background state and studying the linear stability of the small fluctuations around such a state. Weakly non-linear theories can be constructed using higher order expansions terms. While these methods have undoubtedly great value for elucidating the relevant physical processes, they are unable to follow the dynamics of a turbulent atmosphere. We provide a first example of extension of the classical stability analysis to a non-linearly evolving quasi-geostrophic flow. The so-called covariant Lyapunov vectors (CLVs) provide a covariant basis describing the directions of exponential expansion and decay of perturbations to the non-linear trajectory of the flow. We use such a formalism to re-examine the basic barotropic and baroclinic processes of the atmosphere with a quasi-geostrophic beta-plane two-layer model in a periodic channel driven by a forced meridional temperature gradient $\Delta T$. We explore three settings of $\Delta T$, representative of relatively weak turbulence, well-developed turbulence, and intermediate conditions.
We construct the Lorenz energy cycle for each CLV describing the energy exchanges with the background state. A positive baroclinic conversion rate is a necessary but not sufficient condition of instability. Barotropic instability is present only for few very unstable CLVs for large values of $\Delta T$. Slowly growing and decaying hydrodynamic Lyapunov modes closely mirror the properties of the background flow. Following classical necessary conditions for barotropic/baroclinic instability, we find a clear relationship between the properties of the eddy fluxes of a CLV and its instability.
CLVs with positive baroclinic conversion seem to form a set of modes for constructing a reduced model of the atmosphere dynamics.
\end{abstract}
\maketitle

%
\section{Introduction}
\label{sec:intro}

A classical topic of dynamical meteorology and climate dynamics is the study of mid-latitude atmospheric variability and the investigation of the unstable eddies responsible for the synoptic weather.
These unstable eddies affect predictability on time scales of the order of a few days and on spatial scales of the order of a few hundreds kilometers \citep{Kalnay2003}.
They play a crucial climatic role of transporting heat poleward, so that their accurate characterization is of utmost importance.
Classical attempts at understanding their properties are based on linearization of some basic state and normal mode analysis, possibly extended to weakly non-linear regimes \citep{Pedlosky1987}, and on the provision of simple climatic closures \citep{Stone1978}, with stochastic models trying to fill in the gap \citep{Farrell1993}.
Two types of energy conversion between the background state and the fluctuations have been proposed. The barotropic instability converts energy between the kinetic energy of the background state and the eddy field.
As a result the momentum gradients in the background profile are reduced by an unstable barotropic process \citep{Kuo1949}.
The second type of instability is related to the presence of a sufficient vertical shear in the background state \citep{Charney1947,Eady1949,Kuo1952}.
The energetics of the so-called baroclinic instability is dominated by the following processes. The available potential energy of the zonal flow is converted into available potential energy of the eddy field, which is then converted into eddy kinetic energy.
As a result of these processes, the center of mass of the atmosphere is lowered and heat is transported against the temperature gradient.
Necessary instability conditions for the linear stability of generic zonal symmetric states are given by the Charney-Stern theorem \citep{Charney1962,Eliassen1983}. 
The baroclinic and barotropic energy conversions between the zonal mean and the eddies underpin the Lorenz energy cycle (LEC), thus providing the link between weather instabilities and climate \citep{Lorenz1955,Lucarini2009,Lucarini2014}, seen as a non-equilibrium steady state.
Simple two layer quasi-geostrophic (QG) models \citep{Pedlosky1964a,Phillips1954a} provide a qualitative correct picture of the synoptic scale instabilities and energetics of the mid-latitude dynamics \citep{Oort1964,Li2007}. 

The classical approach for studying atmospheric variability is based on defining a background state and studying the linear stability of the small fluctuations around such a state. Weakly non-linear theories can be constructed using higher order expansions terms. While these approaches provide useful insight into the mechanisms responsible for instabilities and the non-linear stabilization, they miss the crucial point of allowing for the investigation of the actual properties of the turbulent regimes, where the system evolves with time in a complex manner, and is far from being in the idealized base state considered in the instability analysis \citep{Speranza1988,Hussain1983}.

We would like to approach the problem of studying the instabilities of the atmosphere in a turbulent regime, taking advantage of some recent tools of dynamical systems theory and statistical mechanics, namely the Covariant Lyapunov Vectors (CLVs)\citep{Ginelli2007,Wolfe2007}.
These allow for studying linear perturbations of chaotic atmospheric flows and investigating the dynamics of the tangent space. In the past Lyapunov vectors were proposed as bases to study the growth and decay of linear perturbations and to associate such features to the predictability of the flow and use them in data assimilation, see \citep{Legras1996,Kalnay2003}. 
Ruelle \citep{Ruelle1979} proposed first the idea of a covariant splitting of the tangent linear space (see also \citep{Trevisan1998}. The covariance of this basis is the critical property for a linear stability analysis, since the basis vectors can be seen as actual trajectories of linear perturbations.
The average growth rate of each CLV equals one of the Lyapunov exponents (LE). 
The LEs describe the asymptotic expansion and decay rates of infinitesimal small perturbations of a chaotic trajectory \citep{Eckmann1985}. 
The CLVs provide explicit information about the directions of asymptotic growth and decay in the tangent linear space.
For stationary states the CLVs reduce to the normal modes.
In the case of periodic orbits the CLVs coincide with the Floquet vectors which for example have been obtained for the weakly unstable Pedlosky model \citep{Samelson2001a}.
Samelson also extended this analysis to unstable periodic orbits \citep{Samelson2001b}.

Recently, new methods to compute CLVs for arbitrary chaotic trajectories have been developed by \citet{Ginelli2007}, and \citet{Wolfe2006,Wolfe2007,Wolfe2008}.
These methods allow for computing CLVs for high dimensional chaotic systems and have led to a renewed interest in the related theory. For a comprehensive introduction we refer to  \citet{Kuptsov2012}.
CLVs have been successfully obtained for one and two dimensional systems \citep{Yang2009,Yang2010,Takeuchi2011}. Moreover, they have been studied for simple models of geophysical relevance \citep{Pazo2010,Herrera2011} elucidating the potential benefits of CLVs in Ensemble Prediction Systems over bred vectors and orthogonal Lyapunov vectors.
In this paper, we construct CLVs for the simple two layer QG model introduced by \citet{Phillips1956} and consider three values of the equator-to-pole relaxation temperature difference $\Delta T$, corresponding to low, medium, and high baroclinic forcing, and correspondingly developed turbulence. This is intended as a first step in the direction of studying a hierarchy of more complex models of geophysical flows.

It is of great interest to link the mathematical properties of the various CLVs to their energetics. Thanks to covariance, we are able to construct the Lorenz energy cycle for each CLV, and then deduce the rate of barotropic and baroclinic energy conversion, as well as of frictional dissipation. In this way, we are able to associate the overall asymptotically growing or decaying property of each CLV to specific physical processes.

The main results obtained in this way are the following.
We observe that CLVs with higher LEs gain energy via the
baroclinic conversion, while energy is mainly
lost by friction and diffusion.
This is accompanied by a northward heat transport, while warm air rises in the south and cold air sinks in the north of the channel.
For the lower negative LEs these processes are inverted.
These qualitative features do not depend on $\Delta T$.
As for the barotropic conversion, the fastest growing CLVs (for the two largest values of $\Delta T$) gain energy by transporting momentum away from the baroclinic jet of the background trajectory. For slow growing CLVs and all decaying CLVs the barotropic conversion is always negative. 

Empirical Orthogonal Functions (EOFs) \citep{Peixoto1992} can be used to construct models of reduced complexity \citep{Selten1995,Franzke2005}. Unfortunately, the modes constructed using this approach are based on correlations and are not related to the actual dynamics of the flow.
As a preliminary idea we take a first step towards a comparable concept employing CLVs. We investigate how much variance of the background trajectories can be explained by the CLVs.
We find that CLVs with higher LEs explain the trajectory's variance better than CLVs with lower LEs.
In particular, CLVs with a positive baroclinic conversion can be used as a meaningful reduced basis of the background trajectory for all values of $\Delta T$.

The structure of the paper is the following.  After giving a short overview of the CLVs in section \ref{sec:II}, we explain in \cref{sec:eqdef} the technical details of the model. In \cref{sec:emd} we present a brief summary of the most important aspects of the LEC for our model. 
In \cref{sec:lec} we construct the LEC for each of the CLVs and discuss its relation to the transports of heat and momentum. In \cref{sec:res} the main results are discussed. First the properties of the Lyapunov spectrum are evaluated then the LEC of the CLVs is evaluated and connections are drawn between the background state and the stable and unstable processes described by the CLVs.
Finally, we report on the reconstruction of the variance of the background trajectories using CLVs, comparing the efficiency of using unstable vs. stable modes (\cref{sec:corr}). We conclude the study with a short summary, conclusion and an outlook for future studies. 

\section{Linear Stability Analysis and Lyapunov Vectors}

\label{sec:II}
Before we present the model used in our study (see next \cref{sec:eqdef}), we recapitulate the mathematical background of covariant Lyapunov vectors (CLVs). They form basis of all linear perturbations to a given background trajectory of a dynamical system and fulfill two important properties.
First, they are covariant, hence each element of the basis is a time-dependent solution to the tangent linear equation (see below equation \ref{eq:tanglinear}). This means the basis vectors are linearized approximations of nearby evolving trajectories and we can identify physical processes between them and the background state (see \cref{sec:lec}). Second, the long-term time average of the growth rates of the single CLVs is equal to the Lyapunov exponents. Consequently, they explore the whole tangent space of a dynamical system. 

Let us see these concepts in some detail. We refer to \citet{Kuptsov2012} for a thorough discussion.
We consider covariant linear perturbations of the non-linear autonomous dynamical system
\begin{equation} \label{eq:dyn}
 \ddt \textbf{x} = f(\textbf{x}),
\end{equation}
$\textbf{x} \in \mathds{R}^n$. We perform a linear stability analysis by studying the evolution of an infinitesimal perturbation $\textbf{v}$ to a solution $\textbf{x}_B$ of equation \ref{eq:dyn}. This means, that two assumptions are made. First, we assume $x_B+v$ is a solution of equation \eqref{eq:dyn} and, second, we assume $\textbf{v}$ to be small at all times. This allows for a first order expansion of the tendency equation around $\textbf{x}_B$.
\begin{align} \label{eq:explin}
\begin{split}
 \ddt \left(\textbf{x}_B+\textbf{v}\right) &= f(\textbf{x}_B+\textbf{v})\\
  &\approx f\left(\textbf{x}_B\right) + \sum_i\spardif{f_j}{x_i}\left(\textbf{x}_B\right)\ \textbf{v}_i
\end{split}
\end{align}
Therefore, in the limit, where $\textbf{v}$ is infinitesimal, $\textbf{v}$ obeys the tangent linear equation.
\begin{equation}\label{eq:tanglinear}
  \ddt \textbf{v}_j(t) = \sum_i\spardif{f_j}{x_i}\left(\textbf{x}_B(t)\right)\ \textbf{v}_i(t) =: \sum_i\mathcal{J}_{ji}\left(\textbf{x}_B(t)\right)\ \textbf{v}_i(t).
\end{equation}
Hereby $\mathcal{J}$ is called the tangent linear operator or Jacobian of the system evaluated at $\textbf{x}^B(t)$. Solutions of equation \ref{eq:tanglinear} are nearby trajectories of $\textbf{x}_B$.

We now wish to explain how to construct a covariant basis of n vectors able to span the tangent  space, each associated to a specific Lyapunov exponent.
For this we first define the propagator $\mathcal{F}$ which evolves linear perturbations through time if they are solutions of equation \ref{eq:tanglinear}.
\begin{equation}
 \textbf{v}(t_2)=\mathcal{F}(t_2,t_1)\ \textbf{v}(t_1)
\end{equation}
Considering the asymptotic far future and far past of $\mathcal{F}$, bases of the tangent space can be constructed.
The commonly used orthogonal forward and backward Lyapunov vectors (FLV, BLV) are derived from these limits and are the eigenvectors of the far past operator $W^-(t)$ (BLV) and the far future operator $W^+(t)$ (FLV).

\begin{align}
\begin{split}
W^-\left(t\right)=&\lim_{t'\rightarrow-\infty}\left[\mathcal{F}\left( t,t' \right)^{-T} \mathcal{F}\left( t,t' \right)^{-1} \right]^{\frac{1}{2(t-t')}} \\
W^+\left(t\right)=&\lim_{t'\rightarrow+\infty}\left[\mathcal{F}\left( t',t \right)^T \mathcal{F}\left( t',t \right) \right]^{\frac{1}{2(t'-t)}}
\end{split}
\end{align}

The eigenvectors of the far past operator are the BLV $B_j(t)$ and the eigenvalues are given in the form  $e^{-\lambda_j}$, which means they have an asymptotic growth rate of $\lambda_j$ in the interval $[-\infty,t]$.
For the sake of simplicity we assume the multiplicity of all eigenvalues to be one.
The $\lambda_j$ are the LEs in descending order.
We can span the so called backward subspaces in the following way $V^-_j(t)=span\{B_1(t),...,B_j(t)\}$ \citep{Osedelec1968}. The meaning of the BLV can be understood by the following examples.
Starting at a time t at ''minus infinity'' with an arbitrary phase space volume $dv_k(t=-\infty)$ of dimension $k$, the final phase space volume at time t, the volume $dv_k(t)$, is a subset of the backward subspace $V^-_k(t)$. The average growth rate of the volume is proportional to the sum of the first k Lyapunov exponents.
\begin{align}\label{eq:blv_vol}
\begin{split}
 &\lim_{t'\rightarrow -\infty}\mathcal{F}(t,t') dv_k(t') \subset V^-_k(t)\\
 &\lim_{t'\rightarrow -\infty}\text{vol}(\mathcal{F}(t,t')dv_k(t'))\propto e^{\sum^{k} _{j=1}\lambda_j \cdot (t-t')}
\end{split}
\end{align} 
For the far future operator the relations are similar. The new eigenvectors $F_j$ are the FLV and span the forward subspaces $V_j^+$. We have $V_j^+=span\{F_{n-j+1},...,F_n\}$ and the eigenvalues are $e^{\lambda_{n-j+1}}$ which means they have an asymptotic growth rate of $\lambda_{n-j+1}$ in the interval $[t,\infty]$. The interpretation is similar to the BLV.
\begin{align}\label{eq:clv_vol}
\begin{split}
  &\lim_{t'\rightarrow +\infty}\mathcal{F}(t',t)^{-1} dv_k(t') \subset V^+_{n-k+1}(t)\\
  &\lim_{t'\rightarrow +\infty}\text{vol}(\mathcal{F}(t',t)^{-1}dv_k(t'))\propto e^{\sum^{k} _{j=1}\lambda_j \cdot (t-t')}
\end{split}
\end{align}

These two kinds of bases are not covariant with equation \ref{eq:tanglinear}. Nevertheless the backward and forward subspaces are covariant.
\begin{equation}\label{eq:cov}
\begin{aligned}
B_j(t_2)&\ne\mathcal{F}(t_2 ,t_1)B_j(t_1)& V^-_j(t_2)=\mathcal{F}(t_2 ,t_1)V^-_j(t_1)\\
F_j(t_2)&\ne\mathcal{F}(t_2 ,t_1)F_j(t_1) & V^+_j(t_2)=\mathcal{F}(t_2 ,t_1)V^+_j(t_1)
\end{aligned}
\end{equation}
Fortunately, we can use them to find a covariant basis (the CLVs). The easiest way to understand this is to rewrite the definition of the forward and backward subspaces by characterizing their elements with respect to their growth rates.

\begin{align}\label{eq:growth}
\begin{split}
V^-_j(t)=&\left\{v\left|\lim_{t'\rightarrow \infty}\frac{1}{t'}log\left(||v(t-t')||\right)\le -\lambda_j\right\}\right. \\
V^+_{n-j+1}(t)=&\left\{v\left|\lim_{t'\rightarrow \infty}\frac{1}{t'}log\left(||v(t+t')||\right)\le +\lambda_{j}\right\}\right.
\end{split}
\end{align}

The cut of covariant subspaces is covariant as well. Therefore, the cut of the subspaces $V^-_j(t)$ and $V^+_{n-j+1}(t)$ contains only vectors which have an asymptotic grow rate of $\lambda_j$ on the interval $[-\infty,\infty]$. Hence, the cut is a one dimensional subspace described by normalized vectors $\left\{\textbf{c}_j\left(t\right)\right\}_{j=1\dots n}$. The vector are the CLV. 

Let us summarize some important properties of the CLVs and explain how we refer to them throughout this study.
Since they are covariant, they cannot in general be orthogonal.
Additionally the covariance forces them to be norm independent.
Note, that the BLV and FLV are dependent on the chosen norm, since they are orthogonal.
We also emphasize that there is a one-to-one relationship between the normalized vector $\textbf{c}_j\left(t\right)$ (the CLV) and a time series of growth rates $\lambda_i(t)$ whose average equal to the jth LE $\lambda_j=\lim_{T\rightarrow \infty}\frac{1}{T}\int_0^T dt \lambda_j(t)$. A solution $\textbf{v}$ of equation \ref{eq:tanglinear} at time $t$ with the initial condition $\textbf{v}(t_0)=\textbf{c}_j(t_0)$ has then the following form.
\begin{equation}\label{eq:clv_ts}
\textbf{v}(t)=e^{\int_{t_0}^t\, dt' \lambda_j(t')}\, \textbf{c}_j(t)
\end{equation}
The normalized CLVs $\textbf{c}_j(t)$ solve the following slightly altered equation.
\begin{equation}\label{eq:norm_clv}
\dot{\textbf{c}}_j(t)=\mathcal{J}(\textbf{x}_B(t))\textbf{c}_j(t)-\lambda_j(t)\textbf{c}_j(t)
\end{equation}

The CLVs are a generalization of the classical normal modes of stationary solutions. It is well known that if $\textbf{x}_B$ is a stationary solution, the covariant Lyapunov vectors are identical to the eigenvectors of the time independent tangent linear operator $\mathcal{J}$ from equation \ref{eq:tanglinear}, see \citet{Wolfe2007}. Hence, they are a appropriate generalization of the classical normal modes of stationary states. For periodic background they coincide with the Floquet vectors \citep{Floquet1883,Samelson2001a,Wolfe2006,Wolfe2008}.

In order to compute the CLVs, we use the algorithm proposed by \citet{Ginelli2007}. A detailed explanation is also found in \citet{Kuptsov2012}.
First, the algorithm computes the backward vectors and the corresponding backward subspaces (with the classical method of Benettin steps \citep{Benettin1980}).
Then we start at the end of the this time series and perform a backward iteration in the tangent space along the trajectory of the forward steps with a new set of random initial vectors.
The vectors are chosen, so that each of them lies in one of the backward spaces.
Additionally, the coordinate system of the tangent space is changed to the backward vector basis $B_j(t)$. In this basis the propagator $\mathcal{F}$ as well as the chosen initial random vectors have a upper triangular shape.
This ensures algebraically that the vectors stay in their initial backward subspaces.
This is necessary, since otherwise any vector would align with the fastest growing direction of the backward dynamics due to the finite computational accuracy.
Also from a mathematical perspective vectors of one backward subspace should always stay in the corresponding subspace (see equation \ref{eq:cov}).
The backward iteration leads to a vector which aligns with the fastest growing vector of the backward dynamics in the respective backward subspaces.
Hence, the backward iterated vector is covariant to equation \ref{eq:tanglinear} and converges towards a vector, with a growth rate equal to the Lyapunov exponents (see equation \ref{eq:growth}) which depends on the initial backward subspace.
These vectors are then the desired CLVs.

\section{The Model}
\label{sec:eqdef}

This is the first time that CLVs are computed for a geophysical model and then used to characterize the properties of linear stability of its chaotic solutions.
As a first step towards a series of more sophisticated models of geophysical flows we choose a two layer model of the mid-latitudes featuring the basic baroclinic and barotropic processes of the atmosphere. These are traditionally described by the decomposition into eddies and zonal mean (eddy-mean decomposition - EMD) and the Lorenz energy cycle (see \cref{sec:emd}).

Our model is a spectral version of the classical model introduced by \citet{Phillips1956}. In the horizontal we have a rectangular domain $\left(x,y \right)\in [0,L_x]\times [0,L_y]$. In the vertical there are five pressure levels ($p_{0.5}=0$ hPa, $p_{1}=250$ hPa, $p_{1.5}=500$ hPa, $p_{2}=750$ hPa, $p_{2.5}=1000$ hPa).

\begin{figure*}
\centering
 \includegraphics[width=\textwidth]{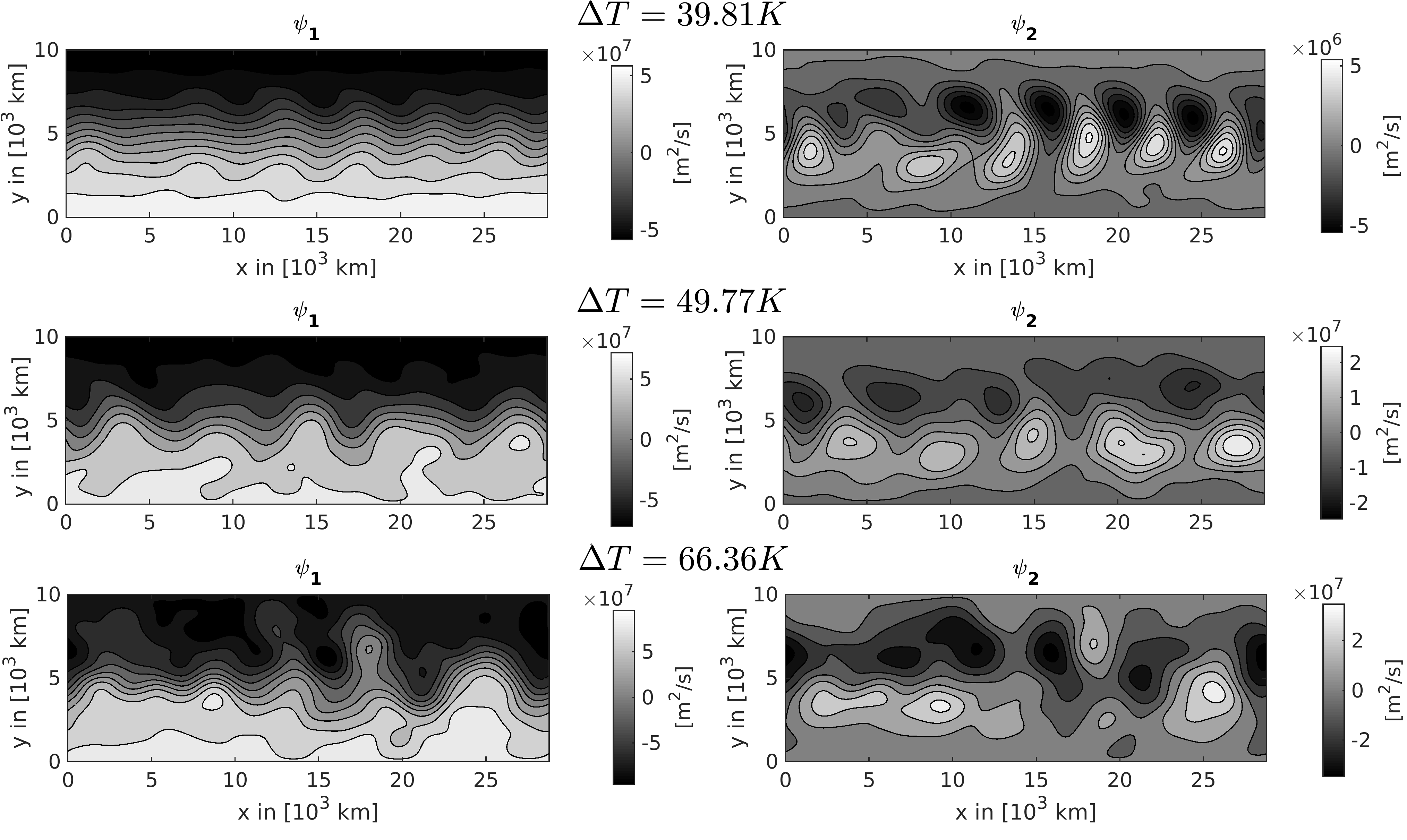}
 \caption{These snapshots show the streamfunction fields for the upper and lower layer. The upper is on the left side and the lower layer is found on the right side. Note, that in order to use a grey scale, the scale of the colorbars is different in every panel. The flow shows a clearly baroclinically unstable and barotropically stable configuration (see \cref{sec:emd}) }\label{fig:snap}
\end{figure*}

In this discretization the hydrostatic equation also gives a simple expression for the temperature $$T_{{1.5}}=\frac{f_0}{R}(\psi_1-\psi_2)=\frac{2f_0}{R}\psi_T.$$ The evolution equations are expressed in form of a partial differential equations (PDE) and are given by
\begin{subequations}
\begin{align}
\begin{split}\label{eq:12a}
\ddt \Delta\psi_{1} =&-\textbf{V}_{1} \cdot \nabla \left(\Delta\psi_1 + f_0 + \beta y\right)\\& + f_0\frac{\omega_ {{1.5}}-\omega_ {{0.5}} }{\Delta p}+k_h\Delta^2\psi_1\\
\end{split}\\
\begin{split}\label{eq:12b}
\ddt \Delta\psi_{2} =&-\textbf{V}_{2} \cdot \nabla \left(\Delta\psi_2 + f_0 + \beta y\right) \\&+ f_0\frac{\omega_ {{2.5}}-\omega_ {{1.5}} }{\Delta p}+k_h\Delta^2\psi_2\\
\end{split}\\
\begin{split}\label{eq:12c}
\ddt \psi_T =&-\frac{\textbf{V}_{1}+\textbf{V}_{2}}{2} \cdot \nabla \psi_T+ S_p \omega +\frac{J}{c_p}+\kappa\Delta \psi_T
\end{split}
\end{align}
\end{subequations}
where equations \eqref{eq:12a} and \eqref{eq:12b} describe the dynamics of the vorticity at the pressure levels 1 \& 2 respectively and equation \eqref{eq:12c} describes the evolution of the temperature field. At the top-level ($p_{0.5}$) the vertical velocity $\omega$ is set to zero, at the lowest layer ($p_{2.5}$) we add an Ekman pumping to account for the friction with the boundary layer ($\omega_{2.5}=\frac{\Delta p}{f_0}\ 2r \Delta\psi_2$).
We express the advection in terms of the  Jacobian $$J(A,B)=\spardif{A}{x}\spardif{B}{y}-\spardif{A}{y}\spardif{B}{x}$$ using the geostrophic stream function $\psi$. The geostrophic velocity $\textbf{V}=(u,v)$ is given by $(-\partial_y \psi,\partial_x \psi)$.
The forcing to the models comes from diabatic heating $J$ given by the newtonian cooling term $c_p r_R\left(\psi_e-\psi_T\right)$. $\psi_e$ accounts for diabatic heating and cooling and provides the baroclinic input into the system. We consider $$\psi_e=\frac{ R \Delta T}{4 f_0} cos(\frac{y\pi}{L_y})$$ where $\Delta T$ is the equator-to-pole temperature difference the system is relaxed to by fast processes such as radiation and convection.
As it is well known, if $\Delta T$ is low the stationary solution is stable, while for higher values of $\Delta T$ baroclinic instability kicks in, so that, when increasing $\Delta T$, through various bifurcations, we reach a state of turbulent motion. Of course it is possible to conduct a sensitivity analysis on $\Delta T$ (see e.g.  \citep{Lucarini2007}), but we will focus solely on three scenarios where the stationary solution is unstable and the steady state of the system is turbulent (see \cref{fig:snap} and table \ref{tab:parameters}).
The full equations of motion in terms of the baroclinic field $\psi_T=\frac{1}{2}\left(\psi_1-\psi_2\right)$ and the barotropic field $\psi_M=\frac{1}{2}\left(\psi_1+\psi_2\right)$ have then the following form: 
\begin{align}
\begin{split}
\ddt\Delta\psi_M =&-J(\psi_M,\Delta \psi_M + \beta y) - J(\psi_T,\Delta\psi_T)\\&-r\Delta(\psi_M-\psi_T)+k_h \Delta^2\psi_M \\
\ddt\,\,\Delta\psi_T =&-J(\psi_T,\Delta \psi_M + \beta y) - J(\psi_M,\Delta\psi_T)\\&+ r\Delta(\psi_M-\psi_T)+k_h \Delta^2\psi_T+\frac{f_0}{\Delta p}\omega\\
\ddt\,\, \psi_T\,\, \,\,=&-J(\psi_M,\psi_T) +S\frac{f_0}{\Delta p}\omega + r_R \left(\psi_{Te}-\psi_T\right)\\& + \kappa \Delta \psi_T. 
\end{split}
\end{align}
Note that $S=S_p\frac{R \Delta p}{2f_0^2}$.The choice of parameters is based on \citet{Phillips1956} and \citet{Lucarini2007}. A list of all parameters and their values is given in Table \ref{tab:parameters}.
\begin{table*}
\centering
\begin{tabular}{llllll}
\toprule
Variables, Operators & Symbol &Unit& Scaling & Value of& \\
\& Constants & & &Factor  & Scaling Factor&\\
\midrule
Stream Function & $\psi$  & $m^2 /s$   & $L^2f_0$&$10^{10}/\pi^2$  &\\
Temperature & $T$           &$K$   & $2f_0^2L^2/R$        &$705.97$  &\\
Velocity        & $\textbf{v}$&$m/s$  & $Lf_0$   &$10^3/\pi$  &\\
Laplace Operator & $\Delta$&$1/m^2$  & $1/L^2$   &$\pi^2/L^2$  &\\
Jacobian & $J(\cdot,\cdot)$&$1/m^2$  & $1/L^2$   &$\pi^2/L^2$  &\\
\midrule
\midrule
Parameters & Symbol & Dimensional & Unit & Scaling&Adimensonal\\
&&Value&&Factor&Value\\
\midrule
Forced Meridional & $\Delta T$  & $40 - 66$ & $K$  & $2f_0^2L^2/R$ &$0.0567 - 0.0936$\\
Temperature Gradient &&&&\\
Eddy-Heat Diffusivity & $\kappa$ & $10^5$ &$m^2/s$& $L^2f_0$&$9.8696 \cdot 10^{-5}$\\
Eddy-Momentum Diffusivity & $k_h$ & $10^5$ &$m^2/s$&$L^2f_0$&$ 9.8696 \cdot 10^{-5}$\\
Thermal Damping & $r_R$ & $1.157\cdot 10^{-6}$ &$1/s$&$f_0$& 0.011\\
Ekman Friction & $r$ & $2.2016 \cdot 10^{-6}$ &$1/s$& $f_0$&0.022\\
Stability Parameter & $S$& $3.33 \cdot 10^{11}$ & $m^2$&$L^2$&0.0329\\
Coriolis Parameter & $f_0$ & $10^{-4}$& $1/s$&$f_0$&1\\
Beta & $\beta$ & $1.599 \cdot 10^{-11}$ & $1/(ms)$ &$f_0/L$&0.509\\
Aspect Ratio &$ a $ & $0.6896 $ & 1 & - & 0.6896 \\
Zonal Length & $L_x$ & $2.9\cdot 10^7$ & $m$&L& $\frac{2\pi}{a}$ \\
Meridional Length & $L_y$ & $10^{7}$ & $m$&L&$\pi$\\
Specific Gas Constant &$R$ &$287.06$&$J/(kg K)$&$R/2$&$2$\\
Pressure &$\Delta p$ &$500 hPa$&$N/m^2$&$\Delta p$& 1\\
\bottomrule
\end{tabular}
\caption{Parameters and Variables used in this model and the respective adimensionalization scheme. Not that the scales for time and length are $t=10^4 s = 1/f_0$ and $L=\frac{10^7}{\pi} m$\label{tab:parameters}}
\end{table*}
The domain is periodic in the x-direction. At $y=0,L_y$ the meridional velocity is set to zero, hence $v=\frac{\partial \psi}{\partial x}=0$ (no flux condition). Since we are solving a second order PDE, a second boundary condition is necessary $$\left.\int_0^{L_x} dx \frac{\partial \psi}{\partial y}\right|_{y=0,L_y}=0$$ (ageostrophic boundary condition) \citep{Pedlosky1987}.
The adimensionalization is performed according to table \ref{tab:parameters}. In the following, we will only use the adimensional model equations. The newtonian cooling stream function in the adimensional form is $\psi_{Te}=\frac{\Delta T}{2}cos(y)$.
\begin{align}\label{eq:eom}
\begin{split}
\ddt\Delta\psi_M =&-J(\psi_M,\Delta \psi_M + \beta y) - J(\psi_T,\Delta\psi_T)\\&-r\Delta(\psi_M-\psi_T)+k_h \Delta^2\psi_M \\
\ddt\,\,\Delta\psi_T =&-J(\psi_T,\Delta \psi_M + \beta y) - J(\psi_M,\Delta\psi_T)\\&+ r\Delta(\psi_M-\psi_T)+k_h \Delta^2\psi_T+\omega\\
\ddt\,\, \psi_T\,\, \,\,=&-J(\psi_M,\psi_T) + S \omega \\&+ r_R \left(\frac{1}{2}\Delta T cos(y)-\psi_T\right) + \kappa \Delta \psi_T. 
\end{split}
\end{align}
The integrations are performed in spectral space using a fourth order Runge-Kutta-Scheme with a fixed time step of $1$ ($2.77$ hours).
With the boundary conditions and the adimensionalization, the stream function has the following form in spectral space.
\begin{align}
\begin{split}
 &\displaystyle\psi(x,y,t)= \sum_{k,l=1}^{N_x,\, N_y} \left(\psi^r(k,l,t)\cos\left(a k x\right) \right.\\
 &\left.+\psi^i(k,l,t)\sin\left(a k x\right) \right)\sin\left(l y\right) + \sum_{l=1}^{N_y} \psi^r(0,l,t) \cos\left(l y\right) \\
\end{split}
\end{align}
Where the spectral cutoff is in the zonal direction at $N_x$ and in the meridional direction at $N_y$.
The total dimension of the model phase space is $2N_y(2N_x+1)$. 
We choose $N_x=10$ and $N_y=12$. Therefore, the total dimension of the model phase space is $504$. The meridional resolution is chosen to accurately approximate the Jacobian $J$ in spectral space. Since its spectral representation requires a projection along the meridional direction we need a sufficiently high $N_y$. The zonal resolution is chosen to include at least all classically unstable normal modes for similar setups of two layer QG models \citep{Holton2004}. 
$\Delta T$ will be chosen in order to cover a weak, medium and chaotic behavior of the model ($39.81 K$, $49.77 K$ and $66.36 K$). All the following results are obtained from 80000 (25 years) long time series.

The resolution in chosen studies regarding solutions of a closely related two layer QG models is significantly higher \citep{Wolfe2006,Wolfe2008}. They obtained Floquet vectors, which are a special case of CLVs obtained if the background state is periodic. In our study, we are obtaining the CLVs for aperiodic background states over a much longer time period of ca. 25 years. Hence, due to our lower resolution we will not be able to study the turbulent cascade, but we will be able to study the behavior of the large scale baroclinic and barotropic processes. Clearly, it would be desirable to use higher resolution, but the fundamental aspects we want to emphasize in this study can already be captured with the current setting. This is also suggested by \cref{fig:snap}, which shows the snapshots of the stream functions. The fields show  baroclinically unstable eddies moving eastward which can be seen from the phase shift between the upper and lower layer.
At the same they are barotropically stable, which can be seen from the angle of the stream function iso lines in regard to the zonal flow. We will discuss this classical large scale description of the mid latitude dynamics represented by an eddy field and a zonal mean in \cref{sec:emd}. 

\section{Atmospheric Circulation and the Lorenz Energy Cycle}
\label{sec:emd} 
\begin{figure*}
\centering
 \includegraphics[width=\textwidth]{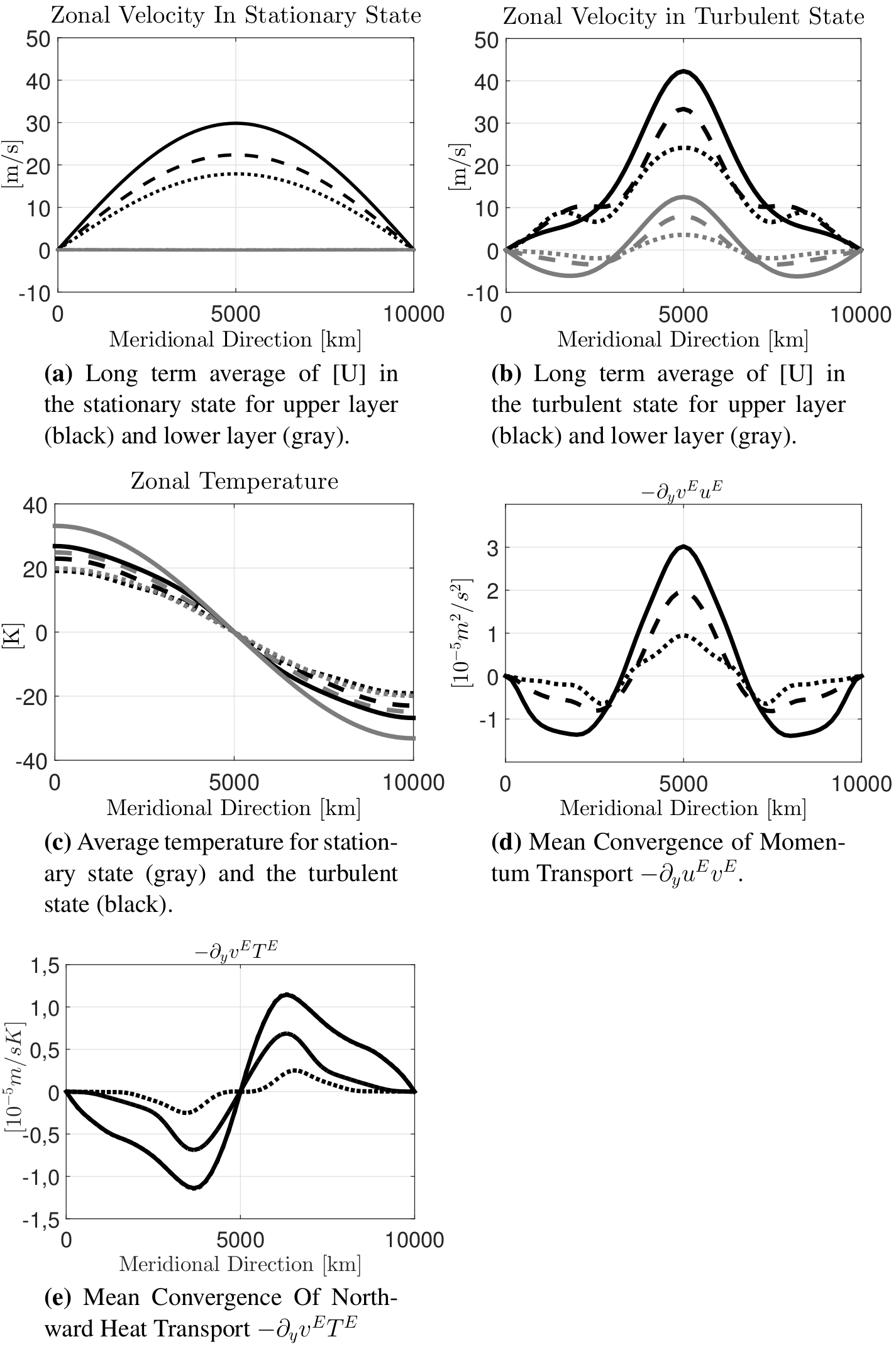}
\caption{The mean state of the trajectory (averaged over a period of 25 years) for the three forced meridional temperature gradients $\Delta T$ (dotted: $39.81 K$ , dashed: $49.77 K$, solid: $66.36 K$). The superscript E indicates the eddy terms.}\label{fig:emd}
\end{figure*}
We recapitulate here the essentials of the classical understanding of the mid-latitudes atmosphere and its turbulent features based on a separation of the trajectory into a zonal mean field ($\left[\psi\right]=\frac{1}{L_x}\int_0^{L_x} dx\, \psi$) and an eddy field ($\psi^E=\psi- \frac{1}{L_x}\int_0^{L_x} dx\, \psi$) \citep{Saltzman1968a} making reference to our model.
We refer to this approach as the eddy-mean decomposition (EMD). 

We study such a decomposition for three different values of the forced meridional temperature gradient $\Delta T$ (dotted: $39.81 K$ , dashed: $49.77 K$, solid: $66.36 K$).
All setups feature an unstable stationary solution and an attractor corresponding to a turbulent solution.

The stationary state is unstable and therefore never observed in the turbulent flow. It is an idealized Hadley equilibrium \citep{Held1980} which  describes a balance of diffusion and newtonian cooling. Recalling the equations of motion (see equation \ref{eq:eom}), the stationary state is the following

\begin{subequations}\label{eq:stat}
\begin{align}
\psi^S_M&=\psi^S_T\cdot\left(\frac{1}{1+\frac{k_h}{r}}\right)\\
\psi^S_T&=\frac{(r_R+\kappa)}{(r_R+\kappa)+S\frac{k_h}{r+k_h}+Sk_h}\psi_{Te}.
\end{align}
\end{subequations}

Here $\Delta T$ is almost equal to the observed temperature difference between the $y=\pi$ and $y=0$ boundary, except for the $\frac{k_h}{r}$ correction term (see figure 9 c).
The upper layer features a broad baroclinic jet, the lower layer features a small easterly flow (see figure 9 b).
In the turbulent solution eddies are transporting heat northward and momentum to the middle of the channel (see figures 9 d, e). This feedbacks on the zonal mean state creating a sharper baroclinic jet with higher velocity gradients than in the stationary state  (see figures 9 a, b, d).
In the lower layer a small eastward jet emerges in the middle of the channel.
North and south of the westerly jet, westward return flows are present in the lower layer balancing the long-term average momentum budget.
The transports do not depend qualitatively on $\Delta T$ and are intensified with a higher $\Delta T$. 

We can further study the turbulence by decomposing the kinetic and potential energy into the zonal mean and eddy contributions. The conversions of energy between these energy reservoirs and the energy losses constitute the Lorenz energy cycle (LEC). Note, that here we use the convention $\int d\sigma=1/(L_xL_y)\int d(x,y) \dots$.
\begin{align}
\begin{split}
 E_{kin}  &=\frac{1}{2}\int d\sigma\, (\nabla \psi_1)^2  + (\nabla\psi_2)^2 \\
&=\overbrace{\frac{1}{2}\int d\sigma\, ([\nabla \psi_1])^2  + \int d\sigma\, ([\nabla\psi_2])^2}^{[E_{kin}]} \\
& +\overbrace{\frac{1}{2}\int d\sigma\, (\nabla \psi^E_1)^2  + \int d\sigma\, (\nabla\psi_2^E)^2}^{E_{kin}^E} \\
E_{pot} &=\frac{1}{S} \int d\sigma\, \psi_T^2\\
&=\underbrace{\frac{1}{S} \int d\sigma\,  (\psi_T^E)^2}_{[E_{pot}]}+\underbrace{\frac{1}{S} \int d\sigma\, \left[\psi_T\right]^2}_{E_{pot}^E}
\end{split}
\end{align}
The LEC is obtained by decomposing the equations of motion into a system of coupled tendency equations for the zonal mean and the eddies \citep{Phillips1956}. The resulting energy conversions and sinks can be then labeled in the following way. \footnote{Z stands for zonal, E for eddy. P is potential energy, K is kinetic energy. NC is newtonian cooling, EF is Ekman friction and KD and HD are kinetic and heat diffusion.}
\begin{align}
\begin{split}
\ddt E_{kin}^E&=C_{ZK\rightarrow EK}+C_{EP\rightarrow EK}+S_{EEF}+S_{EKD}\\
\ddt E_{pot}^E&=C_{ZP\rightarrow EP}-C_{EP\rightarrow EK}+S_{ENC}+S_{EHD}\\
\ddt \left[E_{kin}\right]=&-C_{ZK\rightarrow EK}+C_{ZP\rightarrow ZK}+S_{ZEF}+S_{ZKD}\\
\ddt \left[E_{pot}\right]=&-C_{ZP\rightarrow EP}-C_{ZP\rightarrow ZK}+S_{ZNC}+S_{ZHD}
\end{split}
\end{align} 
\begin{figure*}
\centering
 \includegraphics[width=0.75\textwidth]{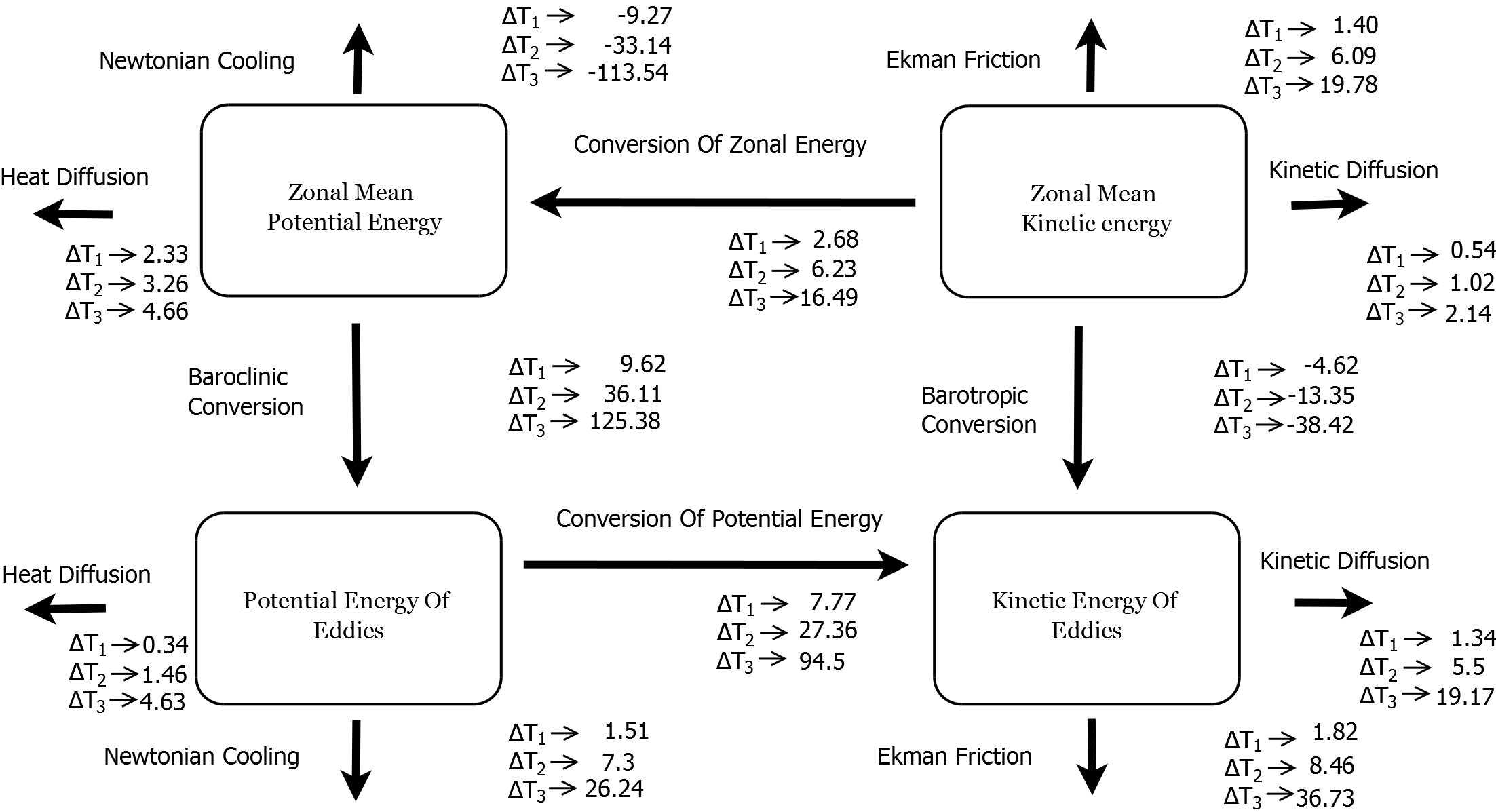}
 \caption{Flow Chart of the Lorenz Energy Cycle for three $\Delta T$ (Units of Conversions are $10^5 m^2/s^3$). The arrows indicate the average sign of the energy conversions, sinks and sources of the zonal and eddy energies. For every temperature gradient ($\Delta T_1=39.81 K$, $\Delta T_2=49.77 K$ and $\Delta T_3=66.36 K$) the dominant source of energy is Newtonian cooling, which inputs energy to the zonal mean potential energy. The important conversions are the baroclinic conversion which is related to the northward heat transport (see \cref{fig:emd} c) and the barotropic conversion related to the center pointed momentum transport (see \cref{fig:emd} d). The main energy losses occour by converting the potential energy of the eddies into kinetic energy, where it is lost mainly due to kinetic diffusion and Ekman friction. We observe an intensification of the cycle for a larger meridional temperature gradient $\Delta T$.\label{fig:lec_dia}}
\end{figure*}
The conversions are the following.
\begin{align}\label{eq:lec_conversion}
\begin{split}
&\text{Baroclinic Conversion }\\
&\mathcal{C}_{ZP\rightarrow EP}= -\int d\sigma\, \frac{2}{S}[v_M^E\psi_T^E]\left[\partial_y\psi_T\right] \\
&\text{Conversion Of Eddy Energy} \\
&\mathcal{C}_{EPEK}=-\int d\sigma\, 2\omega^E\psi_T^E\\
&\text{Conversion of Zonal Energy}\\
&\mathcal{C}_{EK\rightarrow ZK}=-\int d\sigma\, 2 \left[\omega\right]\left[\psi_T\right] \\
&\text{Barotropic Conversion }\\
&\mathcal{C}_{ZPEP}=-\int d\sigma\, \sum_{i=1}^2\left[\partial_yu_i\right]\left(v^E_iu^E_i\right) \end{split}
\end{align}
The long-term averages of these energy conversions are shown in \cref{fig:lec_dia} and are intimately linked to the transports of heat and momentum of the eddies as we will describe in the following.
The baroclinic conversion $\mathcal{C}_{ZP\rightarrow EP}$ quantifies the exchange between the potential energies of the mean and the eddy fields.
In our model the baroclinic conversion is positive, hence the eddies transport heat against the temperature gradient in the zonal mean state (see also figures \ref{fig:emd} c,e). The barotropic conversion $\mathcal{C}_{ZK\rightarrow EK}$ quantifies the exchange between the kinetic energy of the mean and the eddy fields and is negative. This means the eddies transport momentum intensifying velocity gradients in the zonal mean state (see \cref{fig:emd} a, b).
There are conversions from potential to kinetic energy for the zonal mean $\mathcal{C}_{ZP\rightarrow ZK}$ as well as for the eddy field $\mathcal{C}_{EP\rightarrow EK}$. $\mathcal{C}_{EP\rightarrow EK}$ is more relevant and is positive. Hence, on the average warmer air ($\psi_T>0$) rises or colder air ($\psi_T< 0$) sinks lowering the center of mass of the atmosphere and as a result kinetic energy is produced.

There are also sinks of energy related to the various terms of friction, diffusion and newtonian cooling. Note, that all of them are positive sinks of energy whereas the newtonian cooling of the zonal mean is the main input of energy to the LEC.
\begin{equation}\label{eq:sinks}
\begin{aligned}
&\text{Eddy Kinetic Diffusion}&& \text{Eddy Ekman Friction}\\
&\mathcal{S}_{EKD}=-\int d\sigma k_h\sum_{i=1}^{2}\left(\Delta\psi^E_i\right)^2&&\mathcal{S}_{EEF}=\int d\sigma r\psi^E_2\Delta\psi^E_2\\
&\text{Eddy Heat Diffusion}&&\text{Eddy Newtonian Cooling}\\
&\mathcal{S}_{EHD}=\int d\sigma 2 \frac{\kappa}{S} \psi_T^E\Delta\psi_T^E&&\mathcal{S}_{ENC}=-\int d\sigma 2 \frac{r_R}{S} \left(\psi_T^{E}\right)^2\\
&\text{Zonal Kinetic Diffusion}&& \text{Zonal Ekman Friction}\\
&\mathcal{S}_{ZKD}=-\int d\sigma k_h\sum_{i=1}^{2}\left(\partial_y^2\left[\psi_i\right]\right)^2 &&\mathcal{S}_{ZEF}=\int d\sigma r\left[\psi_2\right]\partial_y^2\left[\psi_2\right]\\
&\text{Zonal Heat Diffusion}&&\text{Zonal Newtonian Cooling}\\
&\mathcal{S}_{ZHD}=\int d\sigma 2 \frac{\kappa}{S} \left(\partial_y\left[\psi_T\right]\right)^2&&\mathcal{S}_{ZNC}=-\int d\sigma 2 \frac{r_R}{S}\left[ \psi_T\right]^{2}
\end{aligned}
\end{equation}
Energy is mainly lost by kinetic diffusion and Ekman friction.
The variation of the imposed meridional temperature gradient $\Delta T$ does not change the overall picture of the decomposition into eddy and zonal mean flows. Furthermore, also the heat and momentum transports and the LEC are not changing qualitatively. Overall, eddy energy is gained by baroclinic processes which tends to equilibrate the system, while barotropic processes create a "pointy" jet.

\section{The Lorenz Energy Cycle Of The CLVs}
\label{sec:lec}
Classical stability analysis interprets the growth of the normal modes by introducing a Lorenz energy cycle between the normal modes and the zonal background state (see, e.g. \citep{Holton2004}). We apply this methodology for studying the energy exchange between the CLVs and the turbulent background flow.

For deriving a meaningful definition of a LEC between the background trajectory and the CLVs, it is necessary to bring together the mathematical and the physical perspective on the evolution of the non-linear trajectory and the CLVs. The growth/decay and correlations in the phase space are measured with the help of norms and scalar products. We can connect growth and decay to physical processes, when considering a suitable physical norm. In our model, this role is played by the total energy $E_{tot}$. We can decompose $E_{tot}$ into a sum of kinetic energy $E_{kin}$ and potential energy $E_{pot}$\footnote{ For the averages over the horizontal domain we define $\frac{a}{2\pi^2}\int_0^{2\pi/a} dx\int_0^\pi dy \cdots=\int d\sigma\, ...$. }. 
\begin{align}
  E_{tot}&=\frac{1}{2}\sum_{i=1}^2\int d\sigma\, \left(\nabla\psi_i\right)^2 +\frac{1}{S}\int d\sigma\, \psi_T^2\\
  E_{kin}&=\frac{1}{2}\sum_{i=1}^2\int d\sigma\, \left(\nabla\psi_i\right)^2\\
  E_{pot}&=\frac{1}{S}\int d\sigma\, \psi_T^2
\end{align}
Note that the kinetic energy is in our case a (squared) norm as well, whereas the potential energy is not a norm\footnote{Nevertheless, we will use the potential energy here like a norm and use it to define a "scalar product". This is basically a "correlation like" bilinear form which gives the correlation between two states with respect to the potential energy.}. We can use the norms (meaning the square root of the energies) to define a bilinear scalar product $\left<A,B\right>=\frac{1}{4}\left(\left|\left|A+B\right|\right|^2-\left|\left|A-B\right|\right|^2\right)$.
The average growth rates of the CLVs measured in these or any other norms is equal to the Lyapunov exponents. This can be seen by a simple calculation. Let $||\cdot||$ be an arbitrary norm, $\textbf{c}_j(t)$ the jth CLV and $\lambda_j(t)$ the corresponding time series of the local Lyapunov exponent (for more details see equation \ref{eq:clv_ts}). Then the average growth rate $r$ is given in the following way.
\begin{align}
\begin{split}
r=&\lim_{T\rightarrow\infty }\frac{1}{T}\int_0^T dt \frac{\ddt ||\textbf{c}_j(t)e^{\int_0^t dt' \lambda_j(t')}||}{||\textbf{c}_j(t)e^{\int_0^t dt' \lambda_j(t')}||}\\
=&\lim_{T\rightarrow\infty }\frac{1}{T}\int_0^T dt \,\ddt \log\left(||\textbf{c}_j(t)e^{\int_0^t dt' \lambda_j(t')}||\right)\\
=&\lim_{T\rightarrow\infty }\frac{1}{T} \left\{\log\left(||\textbf{c}_j(T)e^{\int_0^T dt' \lambda_j(t')}||\right)-\log\left(||\textbf{c}_j(0)||\right)\right\}
\end{split}
\end{align} 
Since our model has a finite dimensional phase space, all norms are equivalent. Hence, arbitrary many vectors with the same length in one norm possess a universal finite upper and lower bound for their length in any other norm \citep{MacCluer2009}. Therefore, the norm of $\textbf{c}_j(t)$ is always smaller then a constant $K> 0$: $||\textbf{c}_j||<K$, since its euclidean norm is one. Therefore, the growth rate r can be explicitly calculated.
\begin{equation}
r=\lim_{T\rightarrow\infty }\frac{1}{T}\int_0^T dt \lambda_j(t).
\end{equation}
Hence, the average growth rate of a CLV computed in an arbitrary norm equals always the respective Lyapunov exponent.
Since $E_{tot}=E_{kin}+E_{pot}$, the same is valid for the average growth rate of potential energy ''norm''. In our calculations, we have verified that all three rates equal the doubled Lyapunov exponents given the numerical accuracy. This means that while a CLV is growing or decaying, the ratio of its potential vs kinetic energy is approximately (in a logarithmic sense) constant.

\begin{figure*}
\centering
 \includegraphics[width=\textwidth]{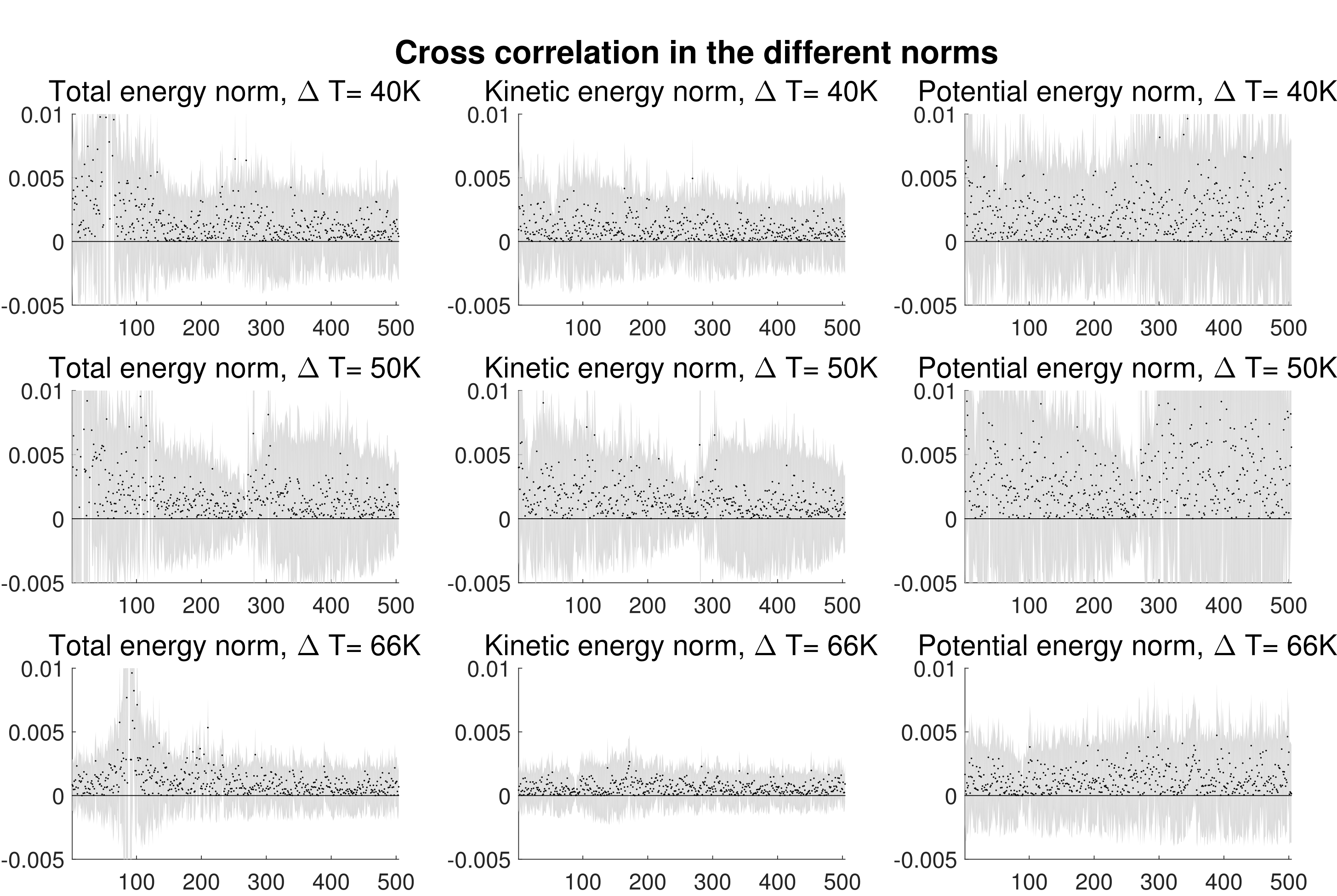}
 \caption{The graphs show the modulus of the average correlation $\frac{<\psi^B, \psi'>}{||\psi^B|| ||\psi'||}$  between the background state and the CLVs, where the bilinear product $<\cdot,\cdot>$ is defined via the kinetic, potential and total energy. The grey shaded areas are the 3 $\sigma$ confidence intervals. We estimated the effective number of degrees of freedom by dividing the time series into blocks corresponding to the e folding time of the autocorrelation function \citep{Leith1973}. The x axis indexes the CLVs. Similar results have been obtained for non zonal stationary states \citep{Niehaus1981}}\label{fig:corr}
\end{figure*}

Let us now consider a linearized solution. Such a solution is the sum of a linear perturbation $(\psi_T',\psi_M')$ constructed with one CLV, the corresponding local growth time series $\lambda(t)$ and the chaotic background solution $(\psi_M^B,\psi_T^B)$  (see \cref{sec:II}).
The energy of this superposition is the sum of the individual energies of the background state and the linear perturbation and an interference term. 
\begin{align}\label{eq:energy}
\begin{split}
 E_{tot} =& E_{kin}+ E_{pot}\\
 E_{kin}=&\sum_i  \int d\sigma\, (\nabla \psi^B_i)^2 + \sum_i 2  \int d\sigma\, \nabla \psi^B_i \nabla \psi'_i \\+&\sum_i   \int d\sigma\, (\nabla \psi'_i)^2\\
  E_{pot}=& \frac{1}{S}  \int d\sigma\, (\psi^B_T)^2+\frac{2}{S} \int d\sigma\, \psi^B_T  \psi'_T\\
+ &\frac{1}{S}  \int d\sigma\, (\psi'_T)^2
\end{split}
\end{align}
We are interested in the long-term average behavior of these terms. A non vanishing long term averaged interference term means that the CLVs evolve ''towards'' the trajectory.
Mathematically, this is equal to a non vanishing average correlation between the linear perturbation and the background state.
This is not the case for any of the three defined energies (see the average correlations in \cref{fig:corr}).
By estimating the effective number degrees of freedom, we can show that with a significance level of 3 $\sigma$ the correlations can be estimated to be zero. 
In fact, this is a necessary prerequisite, since a non vanishing correlation would imply that the background trajectory is not in a non-equilibrium steady state.
This means the trajectory is in a steady state where no growth or decay occurs on average. Note that this is also fulfilled by the CLVs with zero LE. This is expected since the zero growing CLVs are spanned by the tendency $\ddt \psi$ and the meridional velocity $\partial_x \psi$.

What leads linearized solutions to either converge to the background state, to stabilize or to grow infinitely? For this, we must consider the average derivative of energy (by applying $\ddt$ to equation \eqref{eq:energy}). First, the energy of the background state does not grow or decay, hence it cannot contribute to the linearized solution growing or decaying on the average. Second, even tough the interference term is linearly dependent on $\psi'$, it is not contributing to the long term energy growth or decay as well, since we showed that the correlation with the background state vanishes (see \cref{fig:corr}).
Hence, we are left with the time derivative of the third term of equation \eqref{eq:energy} which only depends on the linear perturbation $\psi'$. 
\begin{align}
\begin{split}
\ddt E'_{kin}=&- \int d\sigma\left(  \psi'_1 \ddt  \Delta \psi'_1  + \psi'_2 \ddt  \Delta \psi'_2 \right)\\
\ddt E'_{pot}=&\frac{2}{S} \int d\sigma\left(  \psi'_T \ddt \psi'_T \right)
\end{split}
 \end{align}
From here on we can use the tangent linear equations. The physical interpretation of these terms is then inspired by the classical Lorenz energy cycle used for the EMD picture (see \cref{sec:emd}). 
The equations of motion of the tangent linear model are the following.
\begin{align}
\begin{split}\label{eq:eom_lin}
\ddt\Delta\psi_M' =&-J(\psi_M',\Delta \psi_M^B + \beta y)-J(\psi_M^B,\Delta \psi_M') \\&- J(\psi_T',\Delta\psi_T^B)- J(\psi_T^B,\Delta\psi_T')
\\&-r\Delta(\psi_M'-\psi_T')+k_h \Delta^2\psi_M' \\
\ddt\,\,\Delta\psi_T' =&-J(\psi_T^B,\Delta \psi_M')-J(\psi_T',\Delta \psi_M^B + \beta y)\\& - J(\psi_M^B,\Delta\psi_T')- J(\psi_M',\Delta\psi_T^B)
\\ &+ r\Delta(\psi_M'-\psi_T') +k_h \Delta^2\psi_T'+\omega'\\
\ddt\,\, \psi_T'\,\, \,\,=&-J(\psi_M',\psi_T^B)-J(\psi_M^B,\psi_T') + S \omega' - r_R\psi_T'\\&+ \kappa \Delta \psi_T'. 
\end{split}
\end{align} 
By using the boundary conditions of the model and some algebraic rearrangements we get the following. 
\begin{align}
\begin{split}\label{eq:29}
 \ddt  E'_{kin} =&\int d\sigma\left[\Delta\psi'_1 \textbf{v}_1'\cdot\nabla \psi^B_1 -k_h\left( \psi'_1 \Delta^2\psi'_1\right)\right.\\&\left.+\left< 1 \leftrightarrow 2 \right>- 2 \psi_T' \omega'+ 2 r \psi'_2\Delta\psi'_2\right] 
\end{split}\\
\begin{split}\label{eq:30}
 \ddt  E'_{pot} =&\int d\sigma\left[-\frac{2}{S}\psi_T'\textbf{v}_M'\cdot\nabla \psi^B_T+2\psi_T' \omega'\right.\\&\left.+2 \frac{\kappa}{S} \psi_T'\Delta\psi_T'-2 \frac{r_R}{S} \psi_T^{'2}\right]
 \end{split}
 \end{align}
The interpretation of terms in equation \eqref{eq:29} to \eqref{eq:30} is similar to what was reported in section 4. In section 4, we studied the exchange of energy between the zonal mean state and the eddies. Here, we study the energy conversion between a full non-linear background state and the CLVs.

The conversion from the background potential energy to the perturbation potential energy is the \textit{baroclinic conversion} ($  \mathcal{C}_{BC} $).
\begin{equation}\label{eq:cbc}
 \mathcal{C}_{BC}=\int d\sigma\left[-\frac{2}{S}\psi_T'\textbf{v}_M'\cdot\nabla \psi^B_T\right]
\end{equation}
After applying integration by parts, this term equals the negative correlation between the convergence of heat transport of the CLV ($-\partial_y (v_M'\psi_T')-\partial_x (u_M'\psi_T')$) and the temperature of the background state. Hence, a positive rate means a transport of heat against the temperature gradient in the background state.
The conversion of potential into kinetic energy is described by $\mathcal{C}_{PK}$. 
\begin{equation}\label{eq:cpk}
 \mathcal{C}_{PK}=-2\int d\sigma\, \psi_T' \omega'
\end{equation}
Barotropic processes are contained in the remaining conversion term. This term converts energy to the perturbation kinetic energy (\textit{Barotropic Conversion} $\mathcal{C}_{BT}$). 
\begin{equation}\label{eq:cbt}
\begin{split}
 \mathcal{C}_{BT}=\int d\sigma\left[\Delta\psi'_1 \textbf{v}_1'\cdot\nabla \psi^B_1+\Delta\psi'_2 \textbf{v}'\cdot\nabla \psi^B_2\right]
\end{split}
\end{equation}
We can rewrite this term to explicitly see the connection to the momentum transport of the CLVs and the horizontal divergences in the background flow.
\begin{equation}
\begin{split}
 \mathcal{C}_{BT}=&\int d\sigma\left[-v'_1u'_1\pardif{v^B_1}{x}-v'_1v'_1\pardif{v^B_1}{y}\right.\\&\left.-u'_1u'_1\pardif{u^B_1}{x}-v'_1u'_1\pardif{u^B_1}{y} + \left< 1\leftrightarrow 2\right>\right]
\end{split}
\end{equation}
A positive rate means that the CLVs equilibrate the momentum distribution in the background state.

The other terms are the sinks of the energy cycle due to eddy and heat diffusion, newtonian cooling and Ekman friction and can be correspondingly related to their counterparts in the LEC of the eddies.
\begin{align}
\begin{split}
&\text{Eddy Diffusion}\\
&D_E=\int d\sigma\left[-2k_h\left( \psi'_T \Delta^2\psi'_T+\psi'_P \Delta^2\psi'_P\right)\right]\\
\end{split}\\
\begin{split}\label{eq:ekman}
&\text{Ekman Friction}\\
&F_E=\int d\sigma\ r \psi'_2\Delta\psi'_2\\
\end{split}\\
\begin{split}
&\text{Heat Diffusion}\\
&D_H=\int d\sigma\left[2 \frac{\kappa}{S} \psi_T'\Delta\psi_T'\right]\\
\end{split}\\
\begin{split}
&\text{Newtonian Cooling}\\
&NC=\int d\sigma\left[-2 \frac{r_R}{S} \psi_T^{'2}\right]
\end{split}
\end{align}

Since the CLVs are growing and decaying asymptotically, it will be useful to consider rates instead of time derivatives. Therefore, we normalize all the terms of the LEC by the total energy of the CLVs (at every instant of time). In this way, we will obtain the exponential growth/decay rates of all quantities of the LEC. In general, the mean rate of an energy norm or energy conversion observable $A$ will be $$\lim_{T\rightarrow \infty}1/T\int dt \frac{\ddt A}{E_{tot}}.$$

\section{Physical Properties Of The CLVs}
\label{sec:res}
In this section we present the actual features of the LEC and the associated transports for the CLVs. Additionally, we show the properties of chaoticity that can be derived from the Lyapunov spectrum.

\subsection{The Lyapunov Spectra}\label{sec:LS}
\begin{figure}[hpb]
\centering
 \includegraphics[width=0.7\textwidth]{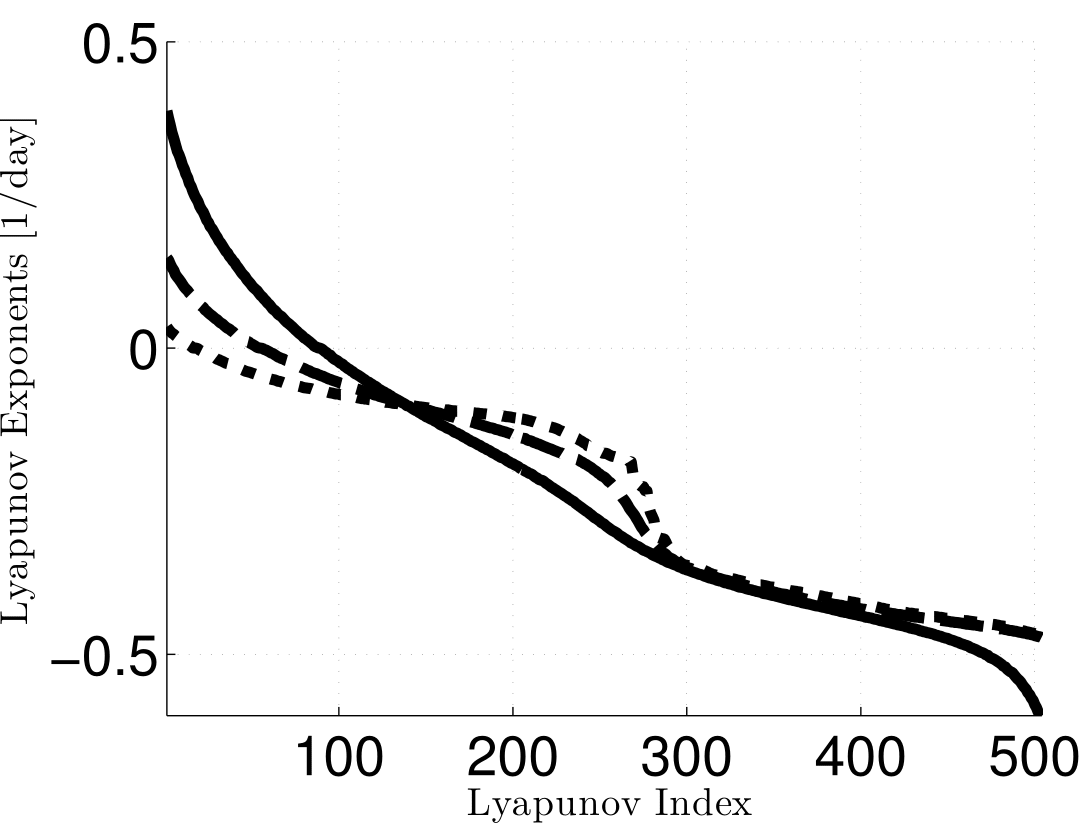}
 \caption{Lyapunov Exponents [1/day] for three meridional temperature gradients  (dotted: $39.81 K$ , dashed: $49.77 K$, solid: $66.36 K$) }\label{fig:lyap}
\end{figure}

The properties of the Lyapunov spectra are presented in this section. All three cases correspond to settings of strong chaos with many positive Lyapunov exponents. This is also reflected in the Kaplan-Yorke dimension and the metric entropy production (see \cref{fig:lyap} and  table \ref{tab:chaos}). The Kaplan-Yorke dimension is defined as
$
D_{KY}= k + \frac{\sum_{i=1}^k \lambda_i}{|\lambda_{k+1}|},
$
where $k$ is chosen, so that the sum of the first $k$ Lyapunov Exponents is positive and the sum of the first $k+1$ is negative.  This dimension is an upper bound of the fractal dimension of the attractor of the system. The metric entropy describes the information creation of the model and is given by the sum of the positive Lyapunov exponents \citep{Eckmann1985}. With increasing $\Delta T$ the Kaplan Yorke dimension and the metric entropy grows monotonically as it was reported by Lucarini et al \citep{Lucarini2007}. While the observed motions are indeed chaotic for the three studied values of $\Delta T$, we can clearly see from these dynamical indicators that turbulence is much better developed for higher value of $\Delta T$.
\begin{table}
\caption{Properties of the attractor}
\label{tab:chaos}
\centering
\begin{tabular}{cccc}
\toprule
$\Delta T$& Positive & Kaplan Yorke & Metric Entropy\\
in [K]& Exponents &  Dimension& in [1/day]\\
\midrule
$39.81$ &$17$& $35.83$& $0.25$\\
$49.77$ &$55$& $125.82$& $3.15$\\
$66.36$ &$88$& $206.80$& $12.51$\\
\bottomrule
\end{tabular}
\end{table}

\subsection{Results for the Lorenz Energy Cycle Of The CLVs}
\label{sec:lecclva}

\subsubsection{Energy Conversion Terms and Sinks}
\label{sec:lecclv}
\begin{figure}
\centering
 \includegraphics[width=\textwidth]{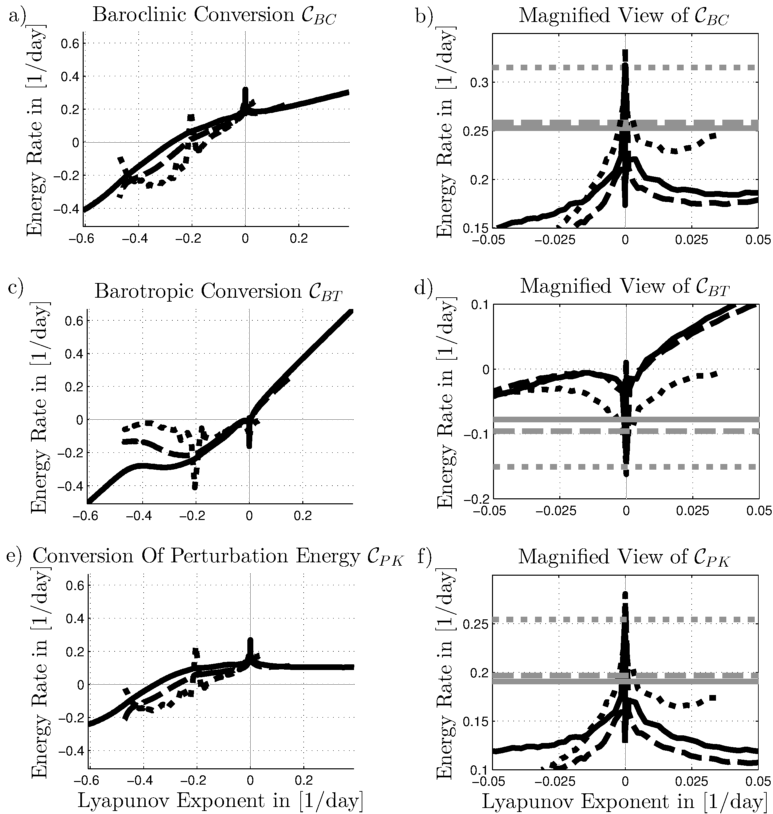}
 \caption{Left Side: The three figures show the dependence of the inputs and the conversion of the Lorenz energy cycle on the corresponding Lyapunov exponent for each of the three meridional temperature gradients  (dotted: $39.81 K$ , dashed: $49.77 K$, solid: $66.36 K$). The magnified view (right side) shows the CLVs with near zero growth rate including the corresponding average eddy observables from the classical Lorenz energy cycle (gray horizontal lines). The y axis units are in 1/day.\label{fig:input_conv}}
\end{figure}
\begin{figure}
\centering
 \includegraphics[width=\textwidth]{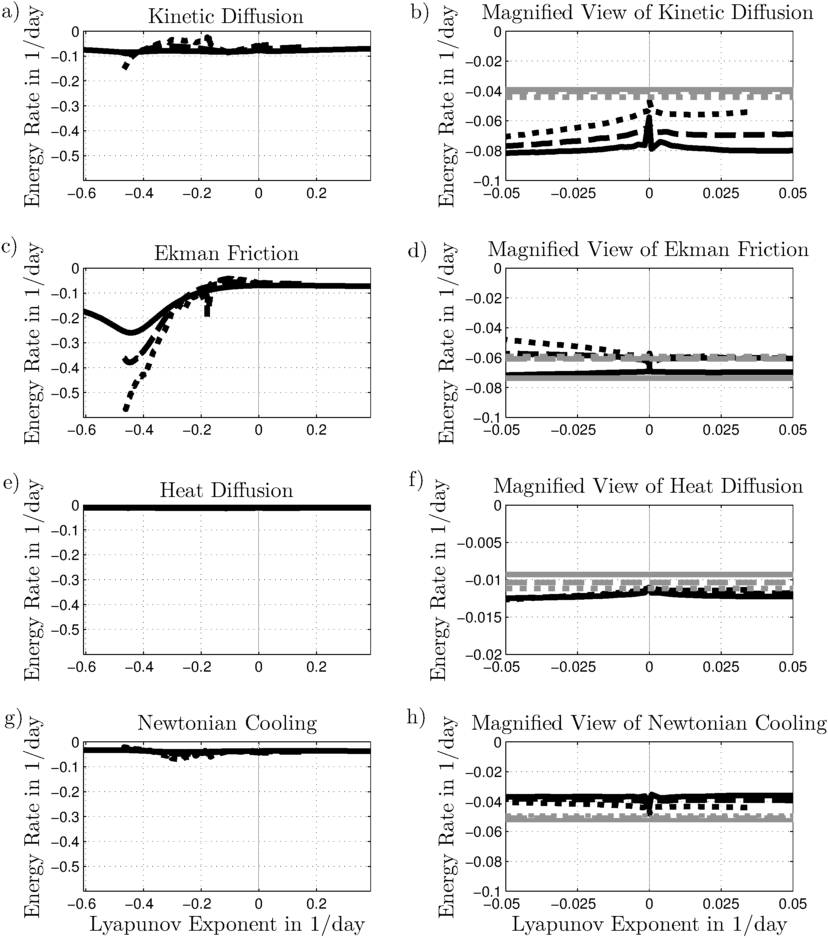}
 \caption{Left Side: The four figures show the dependence of the different sinks of the Lorenz energy cycle on the corresponding Lyapunov exponent for each of the three meridional temperature gradients  (dotted: $39.81 K$ , dashed: $49.77 K$, solid: $66.36 K$). The magnified view (right side) shows the CLVs with near zero growth rate including the corresponding average eddy observables from the classical Lorenz energy cycle (gray horizontal lines). The y axis units are in 1/day. \label{fig:sinks}}
\end{figure}
Now we can unravel the connection of the physics of the CLVs captured by the LEC to their stability properties.
The LEC of the CLVs is given by the long-term averages of the energy budget - normalized to the total energy of the CLV - between a single CLV and the background trajectory (see \cref{sec:lec}). With an abuse of language we will refer to a conversion as ''unstable'' if it is positive and ''stable'' if it is negative.
The barotropic and baroclinic stability properties of the conversions of the LEC vs the value of the corresponding LE for all considered values of $\Delta T$ are shown in \cref{fig:input_conv}.
The baroclinic conversion is positive for roughly half of the CLVs independently of $\Delta T$. This includes all growing CLVs and part of the decaying CLVs.
The barotropic conversion on the other hand is only positive some unstable CLVs in the case of intermediate/large $\Delta T$.
Instead for low $\Delta T$ the barotropic conversion is always negative. 
Hence, with increasing $\Delta T$ the barotropic conversion of all unstable CLVs turns from negative and to positive values for the fast growing CLVs.
Additionally, the conversion of the perturbation energy $\mathcal{C}_{PK}$ follows the sign of the baroclinic conversion $\mathcal{C}_{BC}$. 
We can also observe that being baroclinically unstable is not sufficient to let the CLVs grow because of the effects of friction and diffusion. This result should be considered in respect to findings in more complex models: Here, a high baroclinicity does not always lead to a baroclinic unstable energy growth of the eddies, because a certain threshold of baroclinicty has to be passed \citep{Ambaum2014}. 

Let us now look at the energy sinks of the LEC (see \cref{fig:sinks}). Newtonian cooling, kinetic and heat diffusion show little dependence on the CLVs and $\Delta T$.
The energy loss rate by Ekman friction is large for very stable CLVs (low Lyapunov exponents).
Since the Ekman friction (see equation \eqref{eq:ekman}) is proportional to the kinetic energy of the lower layer of the model, this corresponds to a localization of the flow in the lower layer.
CLVs with the lowest LEs are stable through losses of kinetic energy through $\mathcal{C}_{PK}$ and Ekman Friction.

The slow growing CLVs are of particular interest (see \cref{fig:input_conv}), since they are related to the hydrodynamic Lyapunov modes (HLM) discovered in other non-linear system \citep{Yang2008,Posch2000}. Note that CLVs are superior to orthogonal Lyapunov vectors for finding significant HLMs due to their norm independence \citep{Romero-Bastida2012}. Such CLVs are covariantly evolving solutions of the full non-linear equations which decorrelate very slowly with the background trajectory due to the slow growth rate.
Therefore, it is expected that they might closely represent the properties of the large-scale dominating eddies resulting from removing the mean flow from the actual trajectory of the system.
We compare the slow growing/decaying CLVs with the decomposition of the flow into zonal mean and the eddies (see \cref{sec:emd}) in the magnified view on the right side of the figures \ref{fig:input_conv} and \ref{fig:sinks}. 
They show that the sign and magnitude of conversions from the classical Lorenz energy cycle are comparable to the LEC conversions of these CLVs.
We also see a shift in the behavior of the fast growing CLVs which become barotropically unstable for higher $\Delta T$ and are less comparable to the LEC conversions of the EMD.

We note that the CLVs with corresponding LEs between -1.8 1/day and -3.8 1/day have properties which depend less regularly on the ordering number. This effect results from the fact that these CLVs are quasi-degenerate because of the small difference between the LEs of consecutive CLVs \citep{Kuptsov2012}.

\subsubsection{Convergence of Heat and Momentum Transport and Vertical Velocity}
\label{sec:transport}
\begin{figure*}
\centering
 \includegraphics[width=\textwidth]{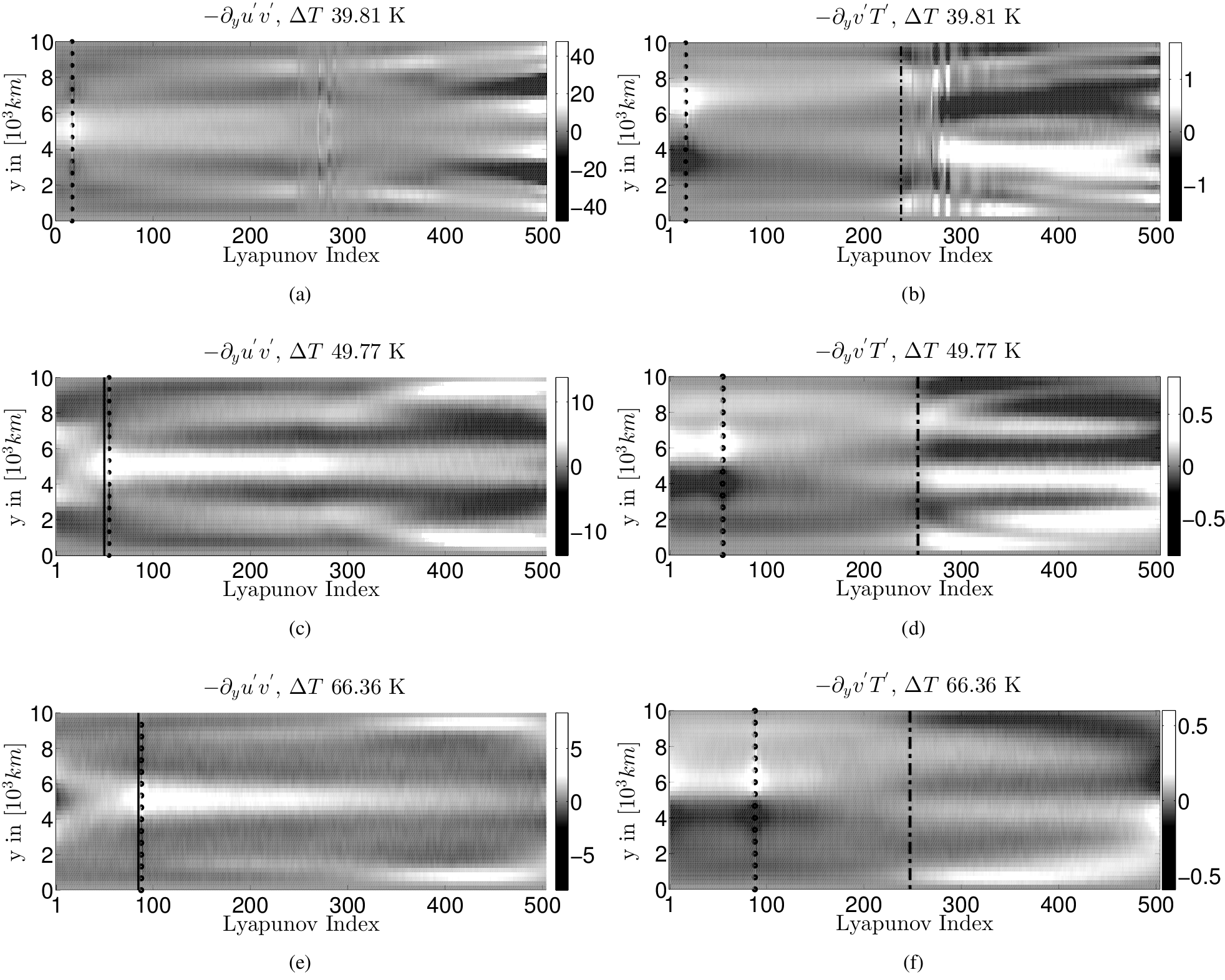}
\caption{Left (a, c, e): The mean zonal profiles of Convergence of Momentum Transport. Right (b, d, f): Northward Heat Transport (b, d, f). 
The x axis indicates the jth CLV.
In (a, c, e) the solid lines indicate the sign switch of the barotropic conversion $\mathcal{C}_{BT}$ from positive to negative.
In (b, d, f) the dash-dotted lines indicate the sign switch of the baroclinic conversion $\mathcal{C}_{BC}$ from positive to negative. The black dotted lines show the CLV with smallest positive LE.
The y axis shows the distribution in the meridional direction in $10^3 km$.\label{fig:transport}}
\end{figure*}

\begin{figure}
\centering
 \includegraphics[width=0.8\textwidth]{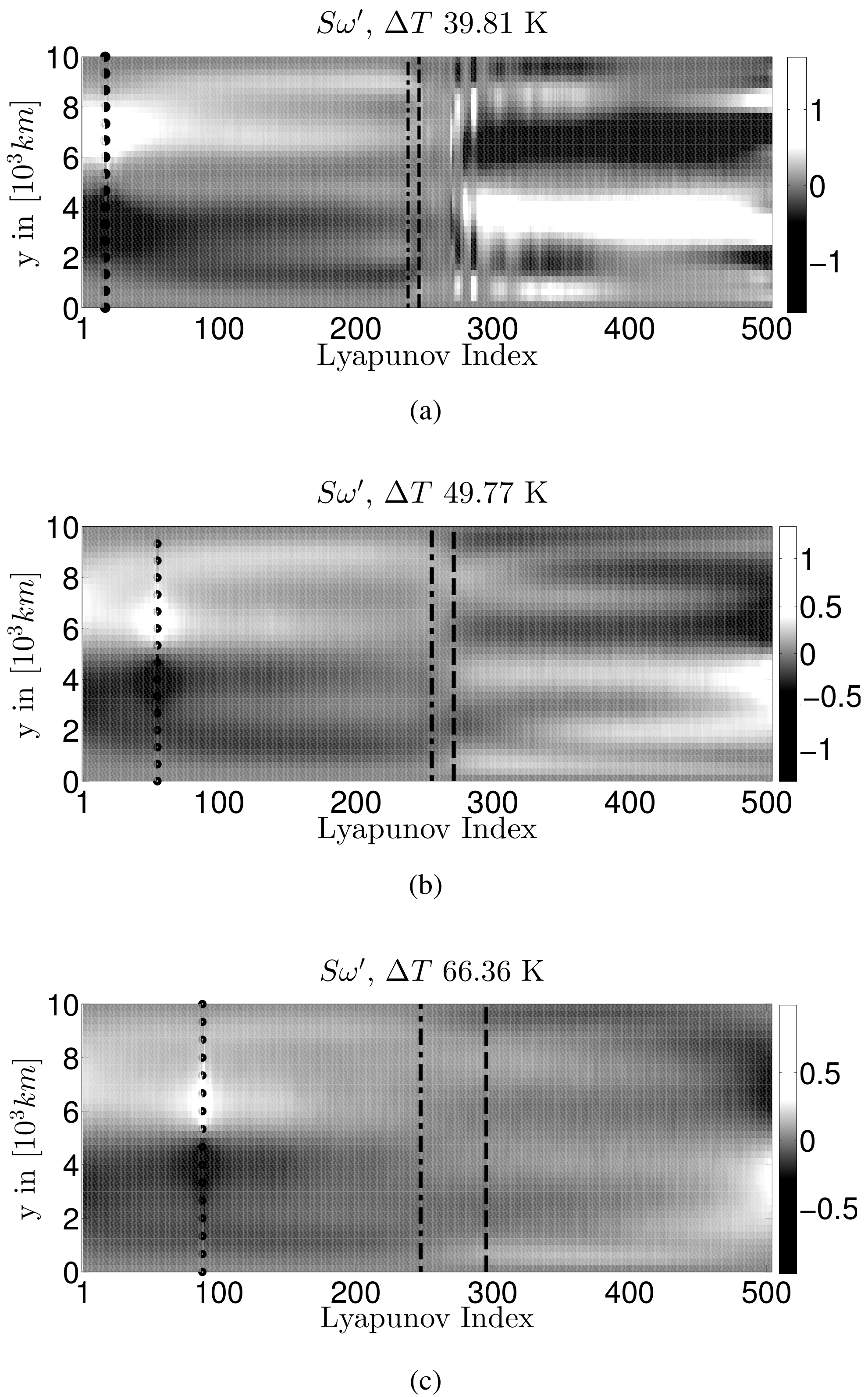}
\caption{The mean zonal profiles of the conversion of heat S? for the three meridional temperature gradients plotted for every CLV ((a) $\Delta T$ = 39.81 K, (b) $\Delta T$ = 49.77 K and (c) $\Delta T$ = 66.36 K). The black vertical dash–dotted
lines indicate the change of sign from positive to negative of the baroclinic conversion CBC. The dashed lines indicate the change of sign in the conversion from
potential to kinetic energy CPK. The black dotted lines show the CLV with the smallest positive LE. The y-axis shows the distribution in the meridional direction in
units of 103 km. The x-axis indicates the jth CLV.\label{fig:vertical}}
\end{figure}

The previously discussed energy conversion ($\mathcal{C}_{BC}$, $\mathcal{C}_{BT}$ and $\mathcal{C}_{PK}$) shown in \cref{fig:input_conv} are intimately related on the heat transport $\textbf{v}_M'T'$, the momentum transport tensor $\textbf{u}'\textbf{v}'$ and the vertical heat transport $S \omega'$ of the CLVs (see equations \ref{eq:cbc}, \ref{eq:cbt} and \ref{eq:cpk}, respectively). Table \ref{tab:transcon} shows systematically the connection between the eddy transports of the CLVs and the gradients of the background state. If the transports and corresponding gradients have a negative correlation then a corresponding conversion is positive. 
If $\mathcal{C}_{BC}$ ($\mathcal{C}_{BT}$) is positive, $\textbf{v}_M'T'$ ($\textbf{u}'\textbf{v}'$) transports heat (momentum) against the gradient of temperature (momentum) in the background state.
If $\mathcal{C}_{PK}$ is positive, warmer air rises and colder air sinks (see \cref{sec:lec}).
We report on the convergence of momentum transport $-\partial_y \sum_i u_i'v_i'$ (see \cref{fig:transport} a,c,e), the convergence of heat transport  $-\partial_y v_M'T'$  (see \cref{fig:transport} b,d,f) and the vertical advection $S \omega'$ (see \cref{fig:vertical} a-c).
\begin{table*}
\centering
\begin{tabular}{cccc}
\toprule
Quantity&Conversions&Transport&Gradients\\
\midrule
Zonal Potential to Eddy Potential Energy &$\mathcal{C}_{BC}$&$\textbf{v}_M'T'$&$\nabla \Psi_T^B$\\
Zonal Kinetic to Eddy Kinetic Energy &$\mathcal{C}_{BT}$&$\textbf{u}'.\textbf{v}'$&$\nabla.\left(u_i^B,v_i^B\right)^T$\\
Eddy Potential to Eddy Kinetic Energy &$\mathcal{C}_{PK}$&$S\omega$ &$\psi_T^B$\\
\bottomrule
\end{tabular}
\caption{Transports and Conversions - The conversion is positive if the correlation between gradient of the background state and eddy transport of the CLVs is negative. Note that for $\mathcal{C}_{PK}$ the baroclinic stream function $\psi_T$ is proportional to the vertical gradient of the stream function.\label{tab:transcon}}
\end{table*}
An integration by parts of equations \ref{eq:cbc} and \ref{eq:cbt} shows that the convergence terms describe the spatial structure of the redistribution of momentum and heat (see \cref{sec:lec}).
We restrict ourselves to zonally averaged quantities, since long-term averages converge to zonally symmetric fields due to the zonal symmetry of the model.

For CLVs with positive baroclinic conversion $\mathcal{C}_{BC}$, we have northward heat transport (see \cref{fig:transport} b,d,f), while a reversed transport is found for CLVs featuring a negative value of $\mathcal{C}_{BC}$..
For lower $\Delta T$ the convergence of heat transport is largest near the center of the channel. As $\Delta T$ increases, the area affected by the transport is extended in the north and south of the channel.
The larger the baroclinic forcing, the more efficient are the CLVs with positive baroclinic conversion in transporting heat northwards, thus reducing substantially the meridional temperature gradient.
Since heat is removed from the very low latitudes and deposited in the very high latitudes, one expects a flattening of the temperature profile.
Note that the near zero CLVs are qualitatively similar, in terms of heat transports, to the most unstable CLVs (see position of the zero LE).

We recall that the barotropic conversion is related to the momentum transport.
As discussed before, for the lowest considered temperature gradient, no CLV features a positive barotropic conversion.
Correspondingly, the momentum transports of the CLVs cause a convergence of momentum in the center of the channel, resulting a pointier jet (see \cref{fig:transport} a).
Things get more complicated when larger values of $\Delta T$ are considered.
In this case, the first CLVs are barotropically unstable, and, in fact, the implied momentum transport (see figs \ref{fig:transport} c,e) of these CLVs causes a depletion on the jet at the center of the channel. This is possible since the horizontal velocity gradients in the background state become sufficiently large (see \cref{fig:emd} b).
For the more turbulent cases, the slow CLVs still feature momentum transport profiles which, instead, support momentum transport towards the center of the channel. This results into the fact that we do observe a pointy jet as mean state of the system (see also \cref{fig:emd} a, b).
Note that most decaying CLVs feature a positive convergence of momentum transport in the middle and on the flanks of the channel. This is connected to the smaller secondary jets on the flanks visible in the mean background state (see \cref{fig:emd} b). 

In figures \ref{fig:vertical} a-c we explore the vertical heat transport $S\omega'$ due to the CLVs. We immediately recognize the signature of baroclinic processes. The CLVs with positive baroclinic conversion feature upward and northward heat transport, while the opposite holds for the CLVs with negative baroclinic conversion rate. This corresponds exactly to the process of release (or creation, in the second case) of available potential energy.

\section{Explaining the variability of the background flow}
\label{sec:corr}
\begin{figure*}
\centering{
 \includegraphics[width=0.95\textwidth]{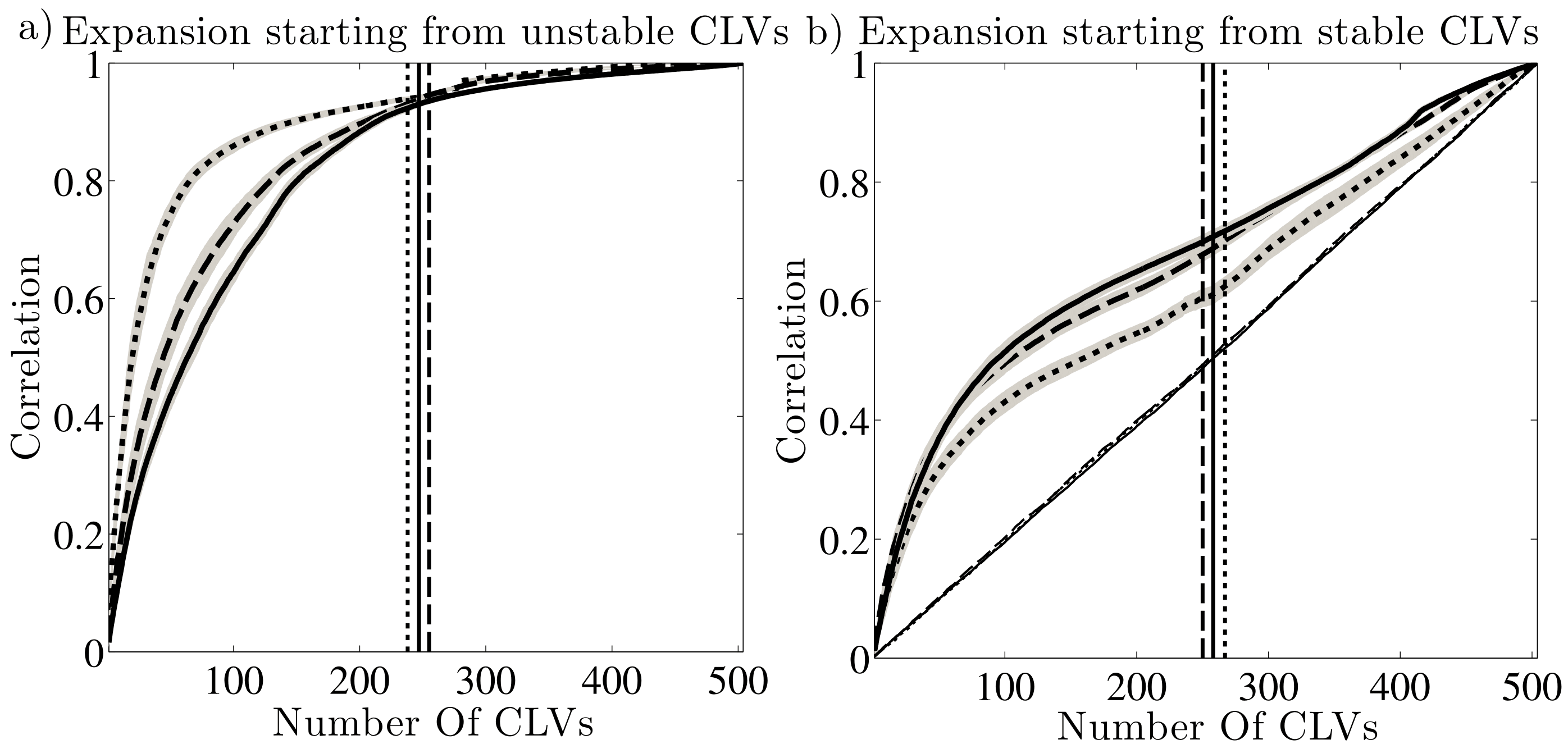}}
 \caption{The panels show the average correlation (solid lines) of the subspaces spanned by the n fastest growing CLVs (a) and the n fastest decaying CLVs (b) with the considered trajectories of $\Delta T$ (dotted: $39.81 K$ , dashed: $49.77 K$, solid: $66.36 K$). The parameter n is indicated on the x axis. The average is done over the mean correlation for different reference points of the CLVs (41 reference points, equally distributed over a 12 years period). The corresponding $\sigma$ area is indicated by the shaded regions. The vertical dashed dotted lines indicate where the expansion includes exactly all baroclinically unstable CLVs (Panel a) or all baroclinically stable CLVs (Panel b). Panel b) also shows the expansion into a randomly chosen basis (almost diagonal dashed lines). The comparison of both panels shows the higher explanatory power of CLVs with a higher LE.\label{fig:correlation}}
\end{figure*}

So far we have studied the linear stability of our model by determining physical properties of CLVs (see \cref{sec:res}). This is equivalent to the physical properties of nearby trajectories to the background trajectory. We can utilize these results to investigate the background state. Their divergent (convergent) evolution is directly linked to the variability of the background state, since the variability in a chaotic solution is caused by the divergence and convergence of nearby trajectories. It is expected that the diverging nearby trajectories dominate the variability over converging nearby trajectories. A classical example for this relation of linear stability and variance can be found interpreting the results of the \citet{Eady1949} model. The linear instability analysis of the Eady model gives as a result modes and corresponding growth/decay rates. The most unstable modes explain qualitatively the variability of the mid latitudes atmosphere to a larger extent. Also, the length scales and growth rates of these modes are comparable to the typical scales of cyclones when reasonable parameters are inserted in the model. Moreover, the Eady modes grow (decay) similar to cyclones due to a vertical westward (eastward) tilt of the troughs and lows which induces a positive (negative) baroclinic energy transfer to the modes from the respective zonal background state and a northward (southward) eddy heat transport. Correspondingly, the CLVs gain (lose) energy by a northward (southward) heat transport $\int d\sigma\, v'T'$ and the positive (negative) baroclinic conversion $C_{BC}=-\frac{2}{S}\int d\sigma\, (\psi_T'\textbf{v}_M'\cdot\nabla \psi^B_T)$. Note that our model features horizontal velocity divergences and therefore the stability of the CLVs is also dependent on the barotropic conversion (see \ref{sec:lec}). 
Having the Eady model in mind, we want to address two questions. First, what is the relation between the Lyapunov exponents and the explained variance of the respective CLVs? Second, how well do baroclinically unstable vs baroclinically stable CLVs explain the variance of the background trajectory? In contrast to other Lyapunov bases of chaotic trajectories the latter question can only be addressed with CLVs due to their covariance.
Note that single CLVs are not correlated with the background trajectory (see \cref{sec:lec}), therefore we will consider subspaces spanned by multiple CLVs. 

We address first how these subspaces of CLVs are constructed. If we consider the CLVs at a reference time $t_R$ as a basis, we have to shift the center of our coordinate system to the corresponding point on the trajectory  ($\textbf{x}_R=(\psi_M(t_R),\psi_T(t_R))$).
Note that the CLVs $\{\textbf{c}_j(t_R)\}$ always form a basis, since they are linear independent by construction.
For our purposes, we choose the following two cascades of sets spanned by CLVs. The first cascade $$\mathcal{B}^{un}_n(t_R)=\operatorname{span}\left\{c_j(t_R)|j=1,\dots,n\right\}$$ contains the CLVs with the n highest LE, the second cascade $$\mathcal{B}^{s}_n(t_R)=\operatorname{span}\left\{c_j(t_R)|j=d-n+1,\dots,d\right\}$$ contains the CLVs with the n lowest LE ($d$ is the total phase space dimension).
The correlation with the subspaces is then defined in the following way.
For each cascade $B^{un/s}_n(t_R)$ the Gram Schmidt algorithm gives an orthogonal basis $\left\{{}^{un/s}\textbf{O}^n_j(t_R)\right\}_{1\le j\le n }$ at the reference point $\textbf{x}_R$.
Hence, the projection of the normalized state vector at time $t$ onto the subspace is
\begin{align}
 \textbf{p}^{un/s}_n(t,t_R)=\sum_{j=1}^n {}^{un/s}\textbf{O}^n_j(t_R) \left<{}^{un/s}\textbf{O}^n_j(t_R),\frac{\textbf{x}(t)}{\left|\left|\textbf{x}(t)\right|\right|}\right>
\end{align}\footnote{The scalar product $\left<\dots,\dots\right>$, is defined in the spectral representation of our model, where $\textbf{x}$ is represented by $(\psi^r_M(k,l),\psi^i_M(k,l),\psi^r_T(k,l),\psi^i_T(k,l))$ and the scalar product of is defined in the following way.
\begin{align*}
&\left<(\psi_M,\psi_T),(\psi'_M,\psi'_T)\right>=\\
&\sum_{l=1}^{N_y}\left\{\left(\sum_{k=1}^{N_x} \psi^r_M(k,l)\psi'^r_M(k,l)+\psi^i_M(k,l)\psi'^i_M(k,l)+\psi^r_T(k,l)\psi'^r_T(k,l)\right.\right.\\
&\left.+\psi^i_T(k,l)\psi'^i_T(k,l) \right) +\psi^r_M(0,l)\psi'^r_M(0,l)+\psi^i_M(0,l)\psi'^i_M(0,l)\\
&\left.+\psi^r_T(0,l)\psi'^r_T(0,l)+\psi^i_T(0,l)\psi'^i_T(0,l) \vphantom{\sum_1^N}\right\}
\end{align*}}.
The correlation for a subspace at the chosen reference point is then
\begin{equation}
\lim_{T\rightarrow \infty}\frac{1}{T}\int_0^T\left<\frac{\textbf{x}(t)}{\left|\left|\textbf{x}(t)\right|\right|}\,,\textbf{p}^{un/s}_n(t,t_R)\right>.
\end{equation}
The average correlation is one if the trajectory lies completely in the subspace defined by $B^{un/s}_n(t_R)$. For the average, we sample the trajectory every 24 hours over a period of 25 years.

\cref{fig:correlation} shows the average correlation as a function of $n$ for the mean and standard deviation of $B^{un/s}_n(t_R)$ obtained from a series of 41 equally distributed reference points over 12 years.
In this figure we also compare the $\mathcal{B}^{s}_n$ cascade with a randomly chosen basis.
For the same n the correlation of $\mathcal{B}^{un}_n$ is always higher than the correlation of $\mathcal{B}^{s}_n$.
Hence, CLVs with higher LEs tend to explain the variance of the background trajectory better than CLVs with lower LEs.
Nevertheless, the expansion into $\mathcal{B}^{s}_n$ is performing better than a random basis with the same size.
This means the explanatory power of the CLVs is related to their stability as we expected it from the Eady model. 
For the second question, we can use the cascades for comparing the baroclinically stable versus the baroclinically unstable CLVs. The baroclinically unstable CLVs are one of the cascades $\mathcal{B}^{un}_n$, whereas the baroclinically stable CLVs are one of the cascades $\mathcal{B}^{s}_n$ (see vertical lines in \cref{fig:correlation}). The expansion into the baroclinically unstable CLVs correlates highly with the trajectory ($\approx 0.94$), whereas the expansion into the baroclinically stable CLVs has a lower correlation with the trajectory ($\approx 0.66$).
The randomly chosen basis of the same size has a correlation of $\approx 0.5$ because its size is about half of the full phase space dimension (see \cref{fig:correlation} b). Hence, the baroclinically unstable CLVs have a significantly higher correlation than the baroclinically stable CLVs. 
Baroclinic instability does not determine the overall stability of the CLVs, but baroclinically unstable CLVs dominate the explanation of the variance of the non linear flow. Moreover, while the Eady modes are rather idealized linear modes, the CLVs are a more general characterization of the flow, since they are trajectories of nearby trajectories.

This allows for suggesting a path for further studies of CLVs. The high quality of the reconstruction variance with the baroclinically unstable CLVs and the weak dependence on $\Delta T$ suggests that this might be a robust way to construct a reduced order model. Traditionally, empirical orthogonal functions of the trajectory are used to
construct such a model of the underlying dynamical system \citep{Holton2004,Franzke2005}. The main limitation of these methods
is that they rely solely on correlations of the trajectory and are
not connected to the equations of motion or to the tangent linear
dynamics which are intimately related to the stable and unstable
processes. Due to their explanatory power and their covariance CLVs could provide a useful tool for further studies in this direction.

\section{Summary \& Conclusion}

Our objective in this study was to determine the physical properties of the tangent linear space of a quasi-geostrophic model of the mid latitudes atmosphere. For this, we made use of new tools (Covariant Lyapunov Vectors )which allow obtaining a covariant basis of this space and allow for investigating linear stability far away from the stationary state of the flow.
Traditional linear stability analysis of the atmosphere investigates normal modes which define the linear stability of typically stationary or zonally symmetric states. This understanding of the dynamics is linked to the decomposition of the atmospheric flow into a zonal mean state and an eddy field.
For our stability analysis of a turbulent background flow we study the evolution of non-linear flows close to a turbulent non-linear background. This is described by a superposition of the background and linear perturbations.
The instabilities and stabilizations of these linear perturbations along the background cannot be reduced to the tangent linear dynamics of a mean profile.
We then study the physical mechanisms responsible
for growth and decay of the small perturbations, by focusing on their energy exchange with the background trajectory using an analysis similar to the traditional Lorenz energy cycle. We investigate baroclinic and barotropic conversion processes and then study the feedbacks.
 
The CLVs provide the appropriate mathematical tool to conduct such an analysis. They span the tangent space in the asymptotic time limit, and they are covariant with the tangent linear dynamics so that they represent actual perturbations to the background trajectory \citep{Ruelle1979}. This allows to examine the link between the stability of the CLVs and their energetic properties given by the energy exchange between the background state and the CLV. We obtain the CLVs with the algorithm proposed by Ginelli et al \citep{Ginelli2007}. 

As a first step towards more sophisticated geophysical models we use a QG two layer model in a periodic channel of the Phillips type \citep{Phillips1956}. It features the basic baroclinic and barotropic processes of the mid-latitudes and is computationally feasible.
Three experiments were conducted with a varying forced meridional temperature gradient $\Delta T$ ($39.81 K$, $49.77 K$, $66.36 K$). These three turbulent regimes feature an increasing Kaplan-Yorke-Dimension and an increasing number of positive LE, so that chaoticity is enhanced. These properties of the LE are consistent with previous findings in a QG model \citep{Lucarini2007}.
All setups feature a baroclinic jet in the upper layer which becomes pointier while increasing $\Delta T$.
We can further characterize the chaotic and turbulent behavior by a decomposition of the flow into a zonal mean and an eddy field.
The Lorenz Energy Cycle of this system features a positive baroclinic conversion accompanied by a northward heat transport and a negative barotropic conversion accompanied by a transport of momentum to middle of the channel. Hence, the turbulent flow reduces the temperature gradients, but intensifies the velocity gradients in the mean state via the meridional transports of the eddies. These processes intensify with increasing $\Delta T$.

The CLVs have a one-to-one relationship to the Lyapunov exponents which are the average growth rates of the CLVs in the euclidean norm. Given that the average growth rates of the CLVs are the same in any norm, a link is provided between physical properties and mathematical properties. Consequently, an energy cycle can be defined between the CLVs and the background flow similar to the classical LEC of the decomposition into eddy field and zonal mean. This allows for connecting baroclinic and barotropic processes and the closely connected heat and momentum transports of the CLVs to their stability properties. 
Roughly half of the CLVs have a positive baroclinic conversion and a northward heat transport against the average temperature gradient of the background state.
As for the barotropic conversion, only fast growing CLVs in the two cases with a higher $\Delta T$ have a positive rate. Hence, they equilibrate the momentum gradients in the background state, if the jet of the background state has a sufficiently high meridional velocity gradient. 
All unstable CLVs have a positive baroclinic conversion, but this is not sufficient for a growing CLV, since friction caused by Ekman pumping and kinetic diffusion counteract this input of energy. The barotropic conversion depends largely on $\Delta T$. Unstable CLVs have a positive barotropic conversion, if the background state features a baroclinic jet with sufficiently large velocity gradients.

We have systematically compared the conversions, sinks and transports of the LEC of the CLVs with the classical LEC obtained by decomposition of the background trajectory into eddy and a zonal mean field. The slowly growing and decaying CLVs exhibit similar properties as they all feature terms of a positive baroclinic conversion and a negative barotropic conversion including the associated momentum transport to the middle of the channel and the northward heat transport.
This is due to the slow decorrelation of the slow growing/decaying CLVs with the background trajectory. In the case of low forcing (low $\Delta T$), we see a correspondence between the classical linear stability analysis \citep{Pedlosky1964a} and our generalized stability analysis with CLVs. In this case even the fastest growing CLVs are slow growing and the mean state is close to the stationary state. Therefore, the most unstable directions exhibit properties similar to the normal modes.

\citet{Posch2000} reported about Hydrodynamic Lyapunov Modes (HLMs), which are very slow growing backward Lyapunov vectors that posses a large scale structure describing macroscopic processes. An unambiguous characterization and detection of HLMs is provided by the CLVs \citep{Romero-Bastida2012}. In general, one expects that in multiscale systems different subsets of CLVs are separated from each other in terms of spatial and temporal scales \citep{Gallavotti2014a}. In our model, we do not find a comparable time and scale separation across the CLVs, because the QG equations are obtained by applying a severe scale analysis to the Navier-Stokes equations, so that only synoptic scales are described.
In fact, analyzing the spectral density of the CLVs, we find that they do not differ in terms of their spatial or temporal scales. Moreover, we find that they have very broad spectral structures in agreement with the fact that the CLVs rather possess heavily localized wave-like features which locally convert and dissipate energy \citep{Herrera2011}. 
The time scales of the CLVs decrease with increasing $\Delta T$ as we expect since larger baroclinicity implies faster growing waves. 
More sophisticated models of the atmosphere should be considered for an analysis with CLVs. In a primitive equations model, we expect to find a clear separation between instabilities related to synoptic processes and instabilities related to mesoscale processes with the latter featuring substantially smaller spatial and temporal scales. Potentially these models would allow for an analysis of HLMs. For parametrizations in such multi-scale systems, following \citet{Gallavotti2014a} one could think of using  the CLVs to filter the model equations for the different scales of motion. Then a parametrization can be derived from the respective statistics of the different scales (similar to \citep{Majda2001}).

We try to ''close the cycle'' and use the CLVs to construct a reduced basis for describing the dynamics of the system, taking into consideration that they describe the unstable and stable modes of variability. This approach differs from EOF-based approaches, because the latter use basis that are only loosely related to the dynamics. Instead, the CLVs are covariant and therefore linked to the dynamics of the turbulent motion. CLVs with positive baroclinic conversion deliver a significantly better explanation of the variance ($\approx 0.94$) then the CLVs with a negative baroclinic conversion ($\approx 0.65$). This is a robust qualitative and quantitative results regardless of the value of $\Delta T$.
Moreover, CLVs with a higher growth rate explain on average the variance of the background trajectory better than a CLVs with a lower growth rate.
This agrees with the general notion that in the end the divergence of nearby trajectories creates the variability found in chaotic models. This reflects e.g. the classical interpretation of the Eady model, where the most unstable linear modes are considered representative of the actual observed variability of the fields.

Future work will deal with two layer QG dynamics with different imposed boundary conditions. This will include on the one hand orography acting on the lower layer \citep{Speranza1985,Charney1979} and on the other hand potential vorticity anomalies imposed on the upper layer. 

Furthermore, the CLVs will be constructed for models featuring time-scale separations like we expect them in the primitive equations and coupled atmosphere/ocean models. We think that a simple coupled ocean-atmosphere models would be an useful first step (e.g. \citep{Vannitsem2013}). 

We will also test the hypothesis whether the CLVs can be used for the construction of a reduced model of the atmospheric circulation. 

It is also promising to study higher resolutions given the discovery of wave-dynamical and damped-advective Floquet vectors by Wolfe and Samelson in a high resolution QG two layer model \citep{Wolfe2006,Wolfe2008}.
Finally, the CLVs should be considered in the context of Ruelle's linear response theory \citep{Ruelle1998a,Lucarini2011,Lucarini2014,Ragone2015}, since they are a basis for the linear response operator of an arbitrary perturbation to a dynamical system.

%

%

\bibliographystyle{apalike}
\bibliography{library}
\end{document}